\documentclass[%
 reprint,
 prl,
 superscriptaddress,
amsmath,aps,amssymb
]{revtex4-1}
\usepackage{dsfont}
\usepackage{graphicx}
\usepackage{dcolumn}
\usepackage{bm}
\usepackage{color}
\usepackage{epsfig}
\usepackage{hyperref}
\usepackage{natbib}
\usepackage{float}
\usepackage{footmisc}
\usepackage{slashed}
\usepackage{comment}
\definecolor{lightred}{rgb}{1,0.5,0.5}
\definecolor{lightgreen}{rgb}{0.5,1,0.5}
\definecolor{lightblue}{rgb}{0.5,0.5,1}
\definecolor{lightcyan}{rgb}{0.5,0.75,0.75}
\definecolor{lightmagenta}{rgb}{0.75,0.5,0.75}
\definecolor{customgreen}{rgb}{0.494,1,0.502}

\newcommand{\eV}{\mathinner{\mathrm{eV}}}
\newcommand{\keV}{\mathinner{\mathrm{keV}}}
\newcommand{\MeV}{\mathinner{\mathrm{MeV}}}
\newcommand{\GeV}{\mathinner{\mathrm{GeV}}}

\definecolor{pinegreen}{rgb}{0.0, 0.47, 0.44}

\begin{document}

\title{Thermal Damping of Neutrino-Coupled Scalar Dark Matter}

\author{Abhishek Banerjee}
\email{abanerj4@umd.edu}
\affiliation{Maryland Center for Fundamental Physics, University of Maryland, College Park, MD 20742, USA}
\author{Ngan H. Nguyen}
\email{nnguye53@jhu.edu}
\affiliation{The William H. Miller III Department of Physics and Astronomy,
The Johns Hopkins University, Baltimore, Maryland, 21218, USA}
\author{Erwin H.~Tanin}
\email{ehtanin@stanford.edu}
\affiliation{Leinweber Institute for Theoretical Physics at Stanford, Department of Physics, Stanford University, Stanford, California 94305, USA}

\begin{abstract}
    We point out that ultralight scalar dark matter that modulates neutrino masses can be significantly thermal damped by cosmic neutrinos in the early universe. This dissipative effect arises as a backreaction from the neutrinos which are being driven slightly out of thermal equilibrium by the scalar. We estimate the rate of such thermal damping and explore its phenomenological implications. For a scalar that is produced early, we find that the effect of thermal damping results in a predictable final abundance largely insensitive to its initial condition while circumventing late time limits. This motivates a parameter-space line to target experimentally.
\end{abstract}
\maketitle

Ultralight bosonic fields can acquire a cosmic abundance and serve as compelling dark matter (DM) candidates~\cite{Essig:2013lka,Ferreira:2020fam,Hui:2016ltb}. 
If they interact with the Standard Model (SM) neutrinos, a wide range of phenomenology ensues, including various signatures in neutrino experiments and altered cosmology~\cite{Berlin:2016woy,Krnjaic:2017zlz,Brdar:2017kbt,Liao:2018byh,Capozzi:2018bps,Huang:2018cwo,Cline:2019seo,Dev:2020kgz, Huang:2021kam,Losada:2021bxx,Chun:2021ief,Dev:2022bae,Huang:2022wmz,Losada:2022uvr,Plestid:2024kyy,Gherghetta:2023myo,Brzeminski:2022rkf}. 
A less explored aspect of such interactions is the dissipative backreaction of slightly non-equilibrium cosmic neutrinos. 
In the early universe, the oscillating DM field continually drives thermal neutrinos away from equilibrium. The neutrinos, in turn, tend to re-thermalize and dissipate the energy it acquired from the DM in a thermodynamically irreversible way. The net effect is a thermal damping of the DM field \cite{Tanin:2017bzm,Mukaida:2012qn,Mukaida:2012bz,Yokoyama:2004pf,Yokoyama:1998ju,Bastero-Gil:2010dgy,Bodeker:2022ihg,Hosoya:1983ke}.

In this \textit{Letter}, we calculate the thermal damping of a scalar DM coupled to neutrinos and point out three implications of this damping \footnote{We discuss thermal damping more generally in a companion paper~\cite{LongPaper}.}. First, thermal damping helps scalars with large initial amplitudes and neutrino-scalar couplings circumvent late-time limits. This makes early-universe scenarios that predict or utilize large-amplitude scalars \cite{Affleck:1984fy,Stewart:1996ai,Dine:1995kz,Tanin:2017bzm,Chang:2025eef,Chang:2024xjd,Chang:2022psj,Banks:1993en,Lyth:1995ka} potentially more viable as these scalars can subsequently be depleted to acceptable levels. Second, the scalar dynamics in the presence of thermal damping by neutrinos has an attractor behavior, where a range of initial conditions converge to the same final abundance, set by the mass and neutrino-coupling of the scalar, making these scenarios predictive. Third, this attractor dynamics implies a coupling-mass relation of the scalar that yields the correct DM abundance. Since this line encapsulates a wide range of initial conditions of the scalar, it serves as an interesting experimental target, we dub \textit{thermal realignment} DM, similar in spirit as thermal misalignment \cite{Batell:2021ofv,Batell:2022qvr,Cyncynates:2024bxw}, and, to lesser degrees, thermal freeze-out \cite{Jungman:1995df} and thermal freeze-in \cite{Hall:2009bx,McDonald:2001vt}.

\paragraph{\textbf{Thermal Damping by Neutrinos.---}}We consider a minimal extension of the SM by coupling the mass eigenstates of the active neutrinos, $\nu_i$ ($i=1,2,3$), to a real scalar field $\phi$ with the following Lagrangian 
\begin{align}
    -\mathcal{L}\supset \frac{1}{2}m_\phi^2\phi^2+\frac{1}{2}(m_{\nu_i}\delta_{ij}-\hat{g}_{ij}g\phi)\nu_i\nu_j+\text{h.c.}\label{eq:Lagrangian}
\end{align}
where we have assumed, for concreteness, that the neutrinos are Majorana. That said, the main part of our analysis occurs entirely in the regime where the neutrinos are relativistic, so the distinction between Dirac (with unexcited right-handed neutrinos) and Majorana is not important here. Furthermore, our main analyses will not depend sensitively on the exact values of the neutrino masses $m_{\nu_i}$ as well as the detailed structure of the Yukawa coupling matrix $\hat{g}_{ij}$ (with the coupling strength $g$ factored out). Unless stated otherwise, we will blur the distinction between the mass eigenstates, effectively setting $m_{\nu_i}\sim m_\nu=0.1\eV$  and $\hat{g}_{ij}=\delta_{ij}$. In the presence of a homogeneous background $\phi$ field that oscillates in time as
\begin{align}
    \phi(t)=\text{Re}\left[\Phi(t)e^{iM_\phi t}\right],
\end{align}
where $M_\phi$ is the effective mass of $\phi$ (to be specified later), $\Phi(t)$ is the oscillation amplitude, and we will assume $|\dot\Phi|/|\Phi|\ll M_\phi$. The neutrinos acquire dynamical mass corrections, such that their effective mass is
\begin{align}
    M_\nu(t)=|m_{\nu}-g\phi(t)|\,.  
\end{align}

\begin{figure*}
    \centering
    \includegraphics[width=0.9\linewidth]{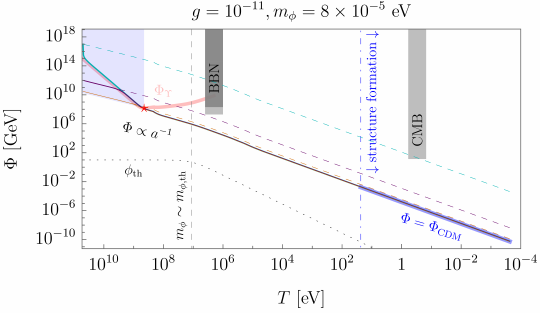}
    \caption{Evolution of the scalar amplitude $\Phi(T)$ as a function of (decreasing) temperature $T$. The thin solid lines are the $\Phi(T)$ for initial conditions such that $\Phi(T=50\GeV)=10^{17}\GeV,\ 10^{12}\GeV,\ 3\times 10^{10}\GeV$, far above the finite-temperature minimum $\phi_{\rm th}$ (dotted line). The dashed lines are the $\Phi(T)$ for the same initial conditions, but with the thermal damping turned off, $\Upsilon=0$. On the $\vee$-shaped pink line labeled $\Phi_\Upsilon$, the oscillation-averaged thermal damping rate $\left<\Upsilon\right>_{\rm osc}$, c.f.~Eq.~\eqref{eq:Upsilonosc}, is equal to Hubble dilution, $\left<\Upsilon\right>_{\rm osc}=2H$. This line stops at $T\approx 2\MeV$, whereupon the neutrinos decouple. The gray prongs represent constraints from BBN (dark gray imposes three relativistic neutrinos, light gray imposes small $\Delta N_{\rm eff}$) and CMB (on the sum of neutrino masses). Here, the scalar-neutrino coupling $g$ and scalar mass $m_\phi$ are chosen such that the $\Phi$ at late times coincides with the amplitude corresponding to the measured DM density today, $\Phi_{\rm CDM}$ (blue band). Initial conditions of $\Phi$ that lie inside the basin of attractor shaded in light blue would track the left side of $\Phi_\Upsilon$, cross the thermal-damping attractor point ($\star$), c.f.~Eq.~\eqref{eq:Apoint}, and converge onto the line that becomes $\Phi_{\rm CDM}$ at late times. The blue dot-dashed line marks the rough point from which the $\phi$ DM must behave as cold DM, with its amplitude scaling as $\Phi\propto a^{-3/2}$, in order to be consistent with large-scale structure observations. In particular, the transition from thermal-mass $m_{\phi,\rm th}$ dominated oscillation ($\Phi\propto a^{-1}$) to bare-mass $m_\phi$ dominated oscillation  ($\Phi\propto a^{-3/2}$), marked with gray dashed line, must occur well before the blue dot-dashed line.}
    \label{fig:PhiT}
\end{figure*}

The modulation of $M_\nu(t)$ leads to the time modulation of the thermal-equilibrium distribution function $f_p^{\rm eq}(t)$ of the neutrinos as
\begin{align}
    f_p^{\rm eq}(t)=\frac{6}{e^{\sqrt{p^2+M_\nu^2(t)}/T}+1}\,, \label{eq:feq}
\end{align}
where we have neglected the  chemical potentials, and the factor of 6 accounts for the spin degrees of freedom of the three mass eigenstates of $\nu_i$. 
This modulation generically causes the actual distribution function $f_p(t)$ to depart from its instantaneous equilibrium value $f_p^{\rm eq}(t)$. 
If the processes that change the number of neutrinos, mainly $\nu\nu\leftrightarrow e^+e^-$, are efficient, the out-of-equilibrium neutrinos will attempt to re-establish thermal equilibrium. This tendency of $f_p(t)$ to chase $f_p^{\rm eq}(t)$ can be captured approximately by the following linearized Boltzmann equation 
\begin{align}
    \delta \dot{f}_p+\Gamma_{\rm th}\delta f_p=\dot{f}_p^{\rm eq},\quad \Gamma_{\rm th}=c_{\rm th} G_F^2T^5\,. \label{eq:Boltzmann}
\end{align}
which is a rewriting of $\dot{f}_p=\Gamma_{\rm th}(f_p^{\rm eq}-f_p)$ in terms of $\delta f_p\equiv f_p-f_p^{\rm eq}$. Here, we have assumed $M_\nu\ll T$ such that $|\delta f_p|\ll f_p^{\rm eq}$, the neutrinos are in a kinetic equilibrium with a temperature $T\ll 100\GeV$, and the relaxation-time approximation~\cite{1954PhRv...94..511B}, where the neutrinos thermalize with a typical weak interaction rate $\Gamma_{\rm th}$ (assumed $\gg H$) \cite{Weinberg:2008zzc}. The prefactor $c_{\rm th}\sim\mathcal{O}(1)$ in Eq.~\eqref{eq:Boltzmann} parametrizes our ignorance of the exact value of $\Gamma_{\rm th}$; note that $c_{\rm th}=1$ reproduces the standard neutrino decoupling temperature, $T\approx 2\MeV$.

In either extremes of $\Phi\ll m_\nu/g$ and $\Phi\gg m_\nu/g$, the $\dot{f}_p^{\rm eq}\propto d(M_\nu^2)/dt$ is sinusoidal with a frequency of $\omega_f=M_\phi$ and $\omega_f=2M_\phi$, respectively. Hence, the steady-state (particular) solution for $\delta f_p$ that solves Eq.~\eqref{eq:Boltzmann} will also be sinusoidal at the same frequency $\omega_f$ \footnote{Transient solutions get damped in a timescale $\Gamma_{\rm th}^{-1}\ll H^{-1}$~\cite{LongPaper}.}. Substituting the ans\"atz $\delta f_p(t)=\text{Re}[\Delta f_p e^{i\omega_f t}]$, where the amplitude $\Delta f_p$ is assumed slowly varying, $|\Delta \dot{f}_p/\Delta f_p|\ll \omega_f$, we find
\begin{align}
    \delta f_p=-\text{Re}\left[\dot{f}_p^{\rm eq}\frac{\Gamma_{\rm th}-i\omega_f}{\Gamma_{\rm th}^2+\omega_f^2}\right].
\end{align}
Assuming the neutrinos are relativistic, we find $\dot{f}_p^{\rm eq}\approx -M_\nu\dot{M}_\nu  f_p^{\rm eq}(1-f_p^{\rm eq}/6)/pT$, and the above equation is \textit{mathematically} equivalent to
\begin{align}
    \delta f_p=-\xi\frac{gM_\nu f_p^{\rm eq}(1-f_p^{\rm eq}/6)\Gamma_{\rm th}}{pT(\Gamma_{\rm th}^2+\omega_f^2)}\left(\dot{\phi}+\frac{\omega_fM_\phi}{\Gamma_{\rm th}}\phi\right)\,, \label{eq:deltaf}
\end{align}
where $\xi=\text{sign}[m_\chi-g\phi]$. The term proportional to $\dot{\phi}$ will lead to thermal damping, regardless of $\xi$, while the term proportional to $\phi$ will lead to a correction to the effective potential of $\phi$.

The non-equilibrium neutrinos will  backreact on the scalar $\phi$ through a source term in the scalar's equation of motion
\begin{align}
    \ddot{\phi}+3H\dot{\phi}+m_\phi^2\phi=\frac{g\xi M_\nu}{2\pi^2}\int p dp\, (f_p^{\rm eq}+\delta f_p), \label{eq:EoM0}
\end{align}
where we have taken the relativistic limit of the expectation value $\left<\nu\nu\right>=\int [d^3p/(2\pi)^3]f_p \xi M_\nu/\sqrt{M_\nu^2+p^2}$~\cite{Peskin:1995ev}. Substituting Eqs.~\eqref{eq:feq} and \eqref{eq:deltaf} into the above, and neglecting the term proportional to $\phi$ in $\delta f_p$, the equation of motion of $\phi$ reduces to the effective form
\begin{align}
    \ddot{\phi}+(3H+\Upsilon )\dot{\phi}+M_\phi^2(\phi-\phi_{\rm th})=0, \label{eq:EoM}
\end{align}
where
\begin{align}
    M_\phi^2&=m_\phi^2+m_{\phi,\rm th}^2,\\
    \phi_{\rm th}&=\frac{m_\nu/g}{1+m_\phi^2/m_{\phi,\rm th}^2},\\
    m_{\phi,\rm th}^2&=\frac{g^2}{2\pi^2}\int p dp\, f_p^{\rm eq}=\frac{g^2}{2}T^2,\\
    \Upsilon&=\frac{1}{\dot{\phi}}\frac{g\xi M_\nu}{2\pi^2}\int p dp\, \delta f_p=\frac{3g^2M_\nu^2}{2\pi^2}\frac{\Gamma_{\rm th}}{\omega_f^2+\Gamma_{\rm th}^2}. \label{eq:Upsilon}
\end{align}
Thus, the neutrinos affect the evolution of $\phi$ in three ways: (1) a shift in the potential-minimum of $\phi$ by $\phi_{\rm th}$ relative to its vacuum value ($\phi=0$), (2) a thermal mass $m_{\phi,\rm th}$, and (3) a thermal damping whose rate is $\Upsilon$. The $\delta f_p$ term in Eq.~\eqref{eq:EoM0} also contributes to $m^2_{\phi,\rm th}$ (through the term proportional to $\phi$ in Eq.~\ref{eq:deltaf}), however this contribution is negligible as we always have $|\delta f_p|\ll f_p^{\rm eq}$. When $\Gamma_{\rm th}\lesssim H$, Hubble expansion (not included in Eq.~\eqref{eq:Boltzmann}) would dominate the evolution of $f_p$ and as a result the thermal damping shuts off, but the minimum shift $\phi_{\rm th}$ and thermal mass $m_{\phi,\rm th}$ remain. In this paper, we are mainly interested in the regime where the scalar amplitude $\Phi$ is much greater than $\phi_{\rm th}$ \footnote{The opposite regime, $\Phi\lesssim \phi_{\rm th}$, corresponds to what is known as the thermal misalignment scenario \cite{Batell:2021ofv,Batell:2022qvr}.}.

The long-term evolution of the scalar amplitude $\Phi$ can be found by multiplying  Eq.~\eqref{eq:EoM} by $\dot{\phi}$ and integrating over time for a half-oscillation of $\phi$, yielding  $\dot{\Phi}/\Phi=-[3H+\left<\Upsilon\right>_{\rm osc}+\dot{M}_\phi/M_\phi]/2$, where the oscillation-averaged thermal-damping rate is defined as $\left<\Upsilon\right>_{\rm osc}\equiv \left[\int_0^{\pi M_\phi^{-1}} dt\,\Upsilon \dot{\phi}^2/(\pi M_\phi^{-1})\right]/(M_\phi^2\Phi^2/2)$, which evaluates to
\begin{align}
 \!\!\!   \left<\Upsilon\right>_{\rm osc}= \frac{3g^2m_\nu^2}{2\pi^2\Gamma_{\rm th}} \begin{cases}
        \displaystyle \frac{\Gamma_{\rm th}^2}{M_\phi^2+\Gamma_{\rm th}^2}\,, &\Phi\ll m_\nu/g\,,\\
        \displaystyle \frac{\Gamma_{\rm th}^2}{4M_\phi^2+\Gamma_{\rm th}^2}\frac{g^2\Phi^2}{4m_\nu^2}\,, &\Phi\gg m_\nu/g\,.
    \end{cases} \label{eq:Upsilonosc}
\end{align}
We plot the evolution of $\Phi$ for a benchmark value of $g=10^{-11}$, $m_\phi=8\times10^{-5}$ eV, and a few representative initial conditions in Fig.~\ref{fig:PhiT}.

\paragraph{\textbf{Evasion of Limits.---}} There are two important cosmological limits on the scenario under consideration. First, the success of Big-Bang Nucleosynthesis (BBN) in predicting the abundances of light-elements requires the presence of three relativistic visible neutrinos and the absence of $\Delta N_{\rm eff}\gtrsim 0.5$ worth of invisible energy density \cite{Pitrou:2018cgg}, at around the temperature of neutrino decoupling. Otherwise, this decoupling temperature, to which the final light-element abundances are sensitive, will be altered considerably. We thus require $g\Phi\lesssim 2\text{ MeV}$ and $M_\phi^2\Phi^2/2\lesssim(7/8)(4/11)^{4/3}\times 0.5\times (2\pi^2/30)T^4$ at $T=2\MeV$. Second, observations of the Cosmic Microwave Background (CMB) damping tail imply that the neutrinos must have free-streamed close to the speed of light at around matter-radiation equality. The analysis of Ref.~\cite{Lorenz:2021alz} on CMB+lensing+BAO+SNe data places the the constraint $\sum_i M_{\nu_i}<0.4\eV$  at $T= 0.3\eV$, which amounts to $3g\Phi|_{T=0.3\,\eV}<0.4\eV$ in our scenario \footnote{The $\sum_i M_{\nu_i}$ is also limited to varying degrees at lower temperatures, however these are less constraining in our scenario, which predicts an $M_\nu$ that is decreasing or constant with time.}. We display these BBN and CMB limits as gray prongs in Fig.~\ref{fig:PhiT}.

We assume in this paper that the $\phi$ field somehow acquired a large amplitude $\Phi\gg\phi_{\rm th}$ from an earlier dynamics, well before BBN. A consequence of thermal damping is that the scalar can have large amplitudes $\Phi$ at early times without spoiling BBN and CMB, or overclosing the universe at late times, if it also has a large coupling $g$.  Thermal damping could ``save" certain combinations of parameters $(g,m_\phi)$ and initial conditions that appears to be ruled out by these limits when thermal damping is not taken into account. This point is illustrated in Fig.~\ref{fig:PhiT}, where we plot several examples of $\Phi$ evolution with and without thermal damping.

\paragraph{\textbf{Thermal-Damping Attractor.---}} If the scalar amplitude is sufficiently large, $\Phi\gg m_\nu/g$ \footnote{We find that in the model under consideration, $\left<\Upsilon\right>_{\rm osc}$ never exceeds Hubble damping in the opposite regime, $\Phi\ll m_\nu/g$. This can be seen as follows. At each temperature $T\lesssim 100\GeV$, there are couplings $g\gtrsim G_F^2T^4$ that saturates the $\left<\Upsilon\right>_{\rm osc}/H$ at its maximum value $\sim G_F^2m_\nu^2T\sim 10^{-10}(T/100\GeV)$. Repeating the same argument at $T\gtrsim 100\GeV$ yields a maximum $\left<\Upsilon\right>_{\rm osc}/H$ that scales as $\propto T^{-3}$. We therefore conclude that in the $\Phi\ll m_\nu/g$ regime, the highest possible $\left<\Upsilon\right>_{\rm osc}/H$ occurs at $T\sim 100\GeV$ and is exceedingly tiny.}, there is an amplitude value $\Phi_\Upsilon(T)$ at each temperature $T$ such that $\left<\Upsilon\right>_{\rm osc}=3H+\dot{m}_{\phi,\rm th}/m_{\phi,\rm th}=2H$ 
\begin{align}
    \Phi_\Upsilon(T)=\frac{4\pi}{g^2\sqrt{3}}(H\Gamma_{\rm th})^{1/2}\left(\frac{4M^2_\phi+\Gamma_{\rm th}^2}{\Gamma_{\rm th}^2}\right)^{1/2}.
\end{align}
This is the $\vee$-shaped thick pink line in Fig.~\ref{fig:PhiT} that bends at around the point where $\Gamma_{\rm th}/M_\phi=\mathcal{O}(1)$. More precisely, this bending occurs at
\begin{align}
    \Phi|_{T=T_*}&= \Phi_*, &&\text{(thermal-damping attractor)} \label{eq:Apoint}
\end{align}
where 
\begin{align}
    T_*&\approx \frac{0.86g^{1/4}}{G_F^{1/2}}=448\MeV\left(\frac{g}{10^{-11}}\right)^{1/4}\label{eq:phistar},\\
    \Phi_*&\approx \frac{18.2}{g^{9/8}M_{\rm pl}^{1/2}G_F^{3/4}}=1.4\times 10^8\GeV\left(\frac{g}{10^{-11}}\right)^{-9/8}.\label{eq:Tstar}
\end{align}
Here, we assume $c_{\rm th}=1$, $H(T_*)=0.3\sqrt{g_*(T_*)}T_*^2/M_{\rm pl}$ with an effective number of relativistic degrees of freedom $g_*(T_*)=61.75$ (not to be confused with the neutrino-scalar coupling $g$) and $M_{\rm pl}=2.4\times 10^{18}\GeV$, neglect the $g_*(T_*)$ dependence of $\Phi_*$ and $T_*$, and assume $g\gg 3\times 10^{-13}(m_\phi/10^{-4}\eV)^{4/5}$ such that $m_{\phi,\rm th}\gg m_\phi$, which is always satisfied for parameters of interest to us. In this parameter space, we have $\Phi_\Upsilon\propto H^{1/2} \Gamma_{\rm th}^{1/2}\propto T^{7/2}$ at $T\gtrsim T_*$ and $\Phi_{\Upsilon}\propto H^{1/2}\Gamma_{\rm th}^{-1/2}m_{\phi,\rm th}\propto T^{-1/2}$, thus explaining the $\vee$ shape of $\Phi_{\Upsilon}(T)$.

The $\Phi_\Upsilon(T)$ bending point, Eq.~\eqref{eq:Apoint}, behaves as an attractor for a range of initial conditions. Above the pink line ($\left<\Upsilon\right>_{\rm osc}\gg 2H$), the amplitude $\Phi$ evolves from an initial value $\Phi_i$ at $t_i$ as $\Phi=\Phi_i/\sqrt{1+\left<\Upsilon\right>_{\rm osc}|_{t_i}(t-t_i)}$. Thus, given our premise, $\Phi$ will reduce to $\Phi_\Upsilon$ in a small fraction of a Hubble time. Below the pink line ($\left<\Upsilon\right>_{\rm osc}\ll 2H$), the amplitude $\Phi$ evolves either as $\Phi\propto a^{-3/2}$ or $\Phi\propto a^{-1}$ for $m_\phi\gg m_{\phi,\rm th}$ or $m_\phi\ll m_{\phi,\rm th}$, respectively \footnote{Since $n_\phi\propto M_\phi\Phi^2\propto a^{-3}$ is an adiabatic invariant (as long as $|\dot{M}_\phi|/M_\phi^2\sim M_\phi H/M_\phi^2\ll 1$), and $M_\phi\propto a^{-1}$ in the thermal-mass dominated case, we must have $\Phi\propto a^{-1}$.}. Therefore, incidentally, $\Phi(T)$ is both steeper than $\Phi_\Upsilon(T)$ above the pink line and less steep than $\Phi_\Upsilon(T)$ below the pink line. This results in an attractor-like behavior for $\Phi(T)$: it tends to approach $\Phi_\Upsilon(T)$ and tracks it. The tracking continues until the turning point, whereupon thermal damping rapidly becomes negligible. Following that $\Phi$ ``freezes out" of $\Phi_\Upsilon$ and continues on an evolution that preserves the adiabatic invariant $n_\phi=M_\phi\Phi^2/2$. 
The bottom line is that a range of initial conditions of $\phi$ converge on the high-temperature part of $\Phi_\Upsilon$, pass through $\Phi_*$ at $T_*$, and eventually lead to the same late-time abundance. 
We dub this as the \textit{thermal realignment} mechanism because thermal damping ``realigns" whatever (sufficiently large) initial misalignment amplitude $\phi$ had to $\Phi_*$ at $T_*$.

We now clarify the range of initial conditions and parameter space that converge to the attractor point. The misalignment of $\phi$ can take place during inflation or at any temperature between $T_*$ and the reheating temperature $T_{\rm reh}$. This could occur, for instance, through de Sitter fluctuations \cite{Tenkanen:2019aij,Kolb:2023ydq,Graham:2018jyp,Alonso-Alvarez:2018tus,Garcia:2023qab} or a spontaneous breaking of a global symmetry for which $\phi$ is a pseudo-Goldstone boson \cite{Preskill:1982cy,Abbott:1982af,Dine:1982ah}. In cases of our interest, $\phi$ begins to oscillate when $M_\phi\sim 3H$, at $T_{\rm osc}\sim 7\times 10^6\GeV(g/10^{-11})$. 
Thus, depending on when the misalignment arises, the scalar could be initially frozen or start oscillating immediately. As previously mentioned, in the parameter space of our interest, incidentally $m_{\phi,\rm th}\gg m_\phi$ at $T\gtrsim T_*$, meaning that $\Phi\propto T$ then. Thus, in this parameter space, any initial conditions such that
\begin{align}
    \Phi(T)\gtrsim \Phi_* \frac{T}{T_*},\quad \text{ at }T_*\lesssim T\lesssim\text{min}[T_{\rm osc},T_{\rm reh}]\label{eq:basin}
\end{align}
will converge and pass through the thermal-damping attractor, Eq.~\eqref{eq:Apoint}.

\begin{figure}
    \centering
    \includegraphics[width=\linewidth]{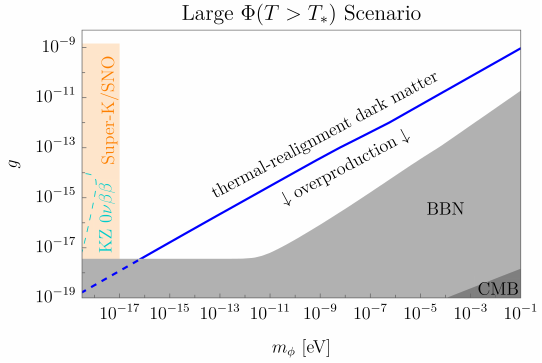}
    \caption{The neutrino-scalar coupling $g$ vs scalar mass $m_\phi$ parameter space for cases with sufficiently large initial amplitudes $\Phi$ to be within the basin of attractor, c.f.~Eq.~\eqref{eq:basin} and the light blue region of Fig.~\ref{fig:PhiT}. On the blue line, the scalar hits the correct DM abundance at late times, c.f.~\eqref{eq:target}. The dashed part of the blue line is inconsistent with cosmic structure formation. The gray shaded regions are constraints from BBN and CMB. The BBN bound includes both the requirements of three relativistic neutrinos and small $\Delta N_{\rm eff}$. These limits are nearly identical and independent of $m_\phi$ if $\phi$ is thermal-mass dominated at $T=2\MeV$, which occurs at $m_\phi\lesssim 10^{-11}\eV$ along the BBN boundary. At higher $m_\phi$, the scalar becomes bare-mass dominated then, and its contribution to $\Delta N_{\rm eff}$ increases with $m_\phi$. It is a mere coincidence that the switching of the blue line from solid to dashed occurs at the boundary of the BBN limit. Also shown are the region ruled out by Super-K/SNO and the ultimate reach of KamLAND-Zen (KZ) $0\nu\beta\beta$ decay experiment as given in Ref.~\cite{Huang:2021kam}, assuming Eq.~\ref{eq:presentabundance} and including a $3\times 10^{5}$ enhancement of the Galactic density relative to the cosmic average.  We stop the plot at $m_\phi=0.1\eV$ because beyond that point $\phi$ stops being wavelike in our Galaxy today.}
    \label{fig:parameterspace}
\end{figure}

\paragraph{\textbf{Thermal-Realignment Dark Matter.---}} Oscillations of the scalar $\phi$ could serve as the full DM \footnote{In a companion paper \cite{LongPaper}, we consider the possibility that $\phi$ constitutes a fraction of the full DM.}. In the absence of $\Upsilon$, the $\phi$ number density $n_\phi=M_\phi \Phi^2/2$ is an adiabatic invariant, which scales as $a^{-3}$. Thus, the present-epoch energy density of $\phi$ can be computed as $\rho_{\phi,0}=M_\phi(T_0) n_\phi(T_*)[s(T_0)/s(T_*)]$, where $s$ is the entropy density of the universe and $T_0$ is the current temperature. We find that its ratio to the current cosmic DM density \cite{Planck:2018vyg} is
\begin{align}
    \frac{\rho_{\phi,0}}{\rho_{\rm DM,0}}=2.9\left(\frac{g}{10^{-11}}\right)^{-7/4}\left(\frac{m_\phi}{10^{-4}\eV}\right)\,,\label{eq:presentabundance}
\end{align}
where we assume $M_\phi\approx m_\phi$ today and take $g_*(T_*)=61.75$. 
Therefore, we have a prediction for the present-day abundance of $\phi$ in terms of the model parameters only. This can also be interpreted as the maximum late-time abundance of $\phi$ that results from a misalignment mechanism taking place before $T_*$, as given in Eq.~\eqref{eq:Tstar}. In addition to hitting the right abundance, in order for $\phi$ to be the DM it needs to facilitate structure formation. The analysis of \cite{Kopp:2018zxp} has placed a limit on the scaling of the energy density of the DM to be matter-like, $|d\ln\rho_\phi/d\ln a+3|/3\lesssim 0.1$, at $z\lesssim 10^{5}$, which roughly translates to the requirement $m_{\phi,\rm th}\lesssim m_\phi$ at redshift $z\sim 10^5$, where the requirement is most restrictive. This amounts to the constraint $g\lesssim 6\times 10^{-6}(m_\phi/10^{-4}\eV)$.

We dub the following $(g,m_\phi)$ line as the \textit{thermal-realignment dark matter} line
\begin{align}
    g\simeq 1.8\times 10^{-11}\left(\frac{m_\phi}{10^{-4}\rm eV}\right)^{4/7}. \label{eq:target}
\end{align}
Along this line, $\phi$ accounts for the full DM abundance for a range of initial conditions within the basin of attractor, Eq.~\eqref{eq:basin}, making it a compelling target to probe experimentally. We now assume that all entries of the Yukawa coupling matrix $\hat{g}_{ij}$ in Eq.~\ref{eq:Lagrangian} has $\mathcal{O}(1)$ values. Existing solar-neutrino-oscillation experiments, such as Super-K and SNO, only probe $m_\phi\lesssim 10^{-17}\eV$ \cite{Berlin:2016woy}, a regime where $\phi$ is ruled out as the DM in our thermal-realignment scenario (although a subcomponent $\phi$ is still viable). DUNE and JUNO can in principle probe higher masses, thanks to their good energy reconstruction, however their sensitivities are far from reaching the target line. For $\phi\nu\nu$ couplings that allow for $\Delta L=2$ transitions, experiments looking for $0\nu\beta\beta$ decay (in this case Bose-enhanced by the $\phi$ background) may have better sensitivities than DUNE and JUNO. For instance, the ultimate sensitivity of KamLAND-Zen as given in Ref.~\cite{Huang:2021kam} can probe close to the target line. Future generations of such an experiment might eventually reach the target line, since, unlike solar-neutrino experiments, the $m_\phi$ reach is not necessarily limited to be $\lesssim 10^{-17}\eV$.

We note that on neither side of the blue line in Fig.~\ref{fig:parameterspace} is $\phi$ ruled out as the DM. At couplings $g$ above the blue line, the $\phi$ DM must be produced at a relatively late time, after the thermal damping becomes insignificant. This implies that the DM production must occur within the temperature window of $\keV\lesssim T\lesssim T_{\rm max}$, where $T_{\rm max}$ is at most $T_*$ but could be as low as 2 MeV. The DM parameter space considered in Refs.~\cite{Berlin:2016woy,Krnjaic:2017zlz,Huang:2021kam}, for instance, lies above our blue line, which implies that those scenarios require a late-time (in the aforementioned sense) DM generation mechanism in order to avoid underproduction of DM due to thermal damping. Below the blue line, the thermal-realignment scenario predicts an overabundance of DM. However, this conclusion only applies for initial conditions within the basin of attractor Eq.~\eqref{eq:basin}. Initial conditions with smaller initial amplitudes $\Phi$ remain viable.

We have also checked that for parameters of our interest, the following is true. The thermal-damping attractor point, Eq.~\eqref{eq:Apoint}, satisfies $T_*\gg 2\MeV$ and $g\Phi_*\lesssim T_*$, which ensure the neutrinos are coupled and relativistic at temperatures where thermal damping is estimated to be most important, and $\Phi_*\lesssim m_{\rm pl}$, which avoids superplanckian scalar amplitude. The approximation $\Phi\gg \phi_{\rm th}$ applies throughout the evolution of $\phi$. Along the thermal-misalignment target, we always have $g\Phi\gg m_\nu$ at $T=2\MeV$, ensuring that $\Phi_\Upsilon$ exists at  $T\gtrsim 2\MeV$. There is no danger of thermalization of $\phi$ quanta via weak-mediated bremsstrahlung $\nu\nu\rightarrow\nu\nu\phi$ (most important at $T\sim 100\GeV$) and via neutrino-to-scalar annihilation $\nu\nu\rightarrow \phi\phi$ (most important at late times). Evaporation of the scalar condensate $\phi$ via neutrino scatterings $\nu\phi\rightarrow\nu\delta\phi$ is inefficient. Non-adiabatic production of neutrinos may occur when $\Phi>m_\nu/g$, however it does not drain the energy of $\phi$ substantially. The loop-induced quartic coupling $\sim g^4/16\pi^2$ does not affect our analysis, in particular $(g^4/16\pi^2) \Phi_*^4\ll M_\phi^2(T_*)\Phi_*^2$. Moreover, at the present epoch, this quartic coupling limits the scalar amplitude through parametric resonance to far smaller than what is needed to spindown a black hole appreciably, thus alleviating the superradiance limit \cite{Lambiase:2025twn,Mathur:2020aqv,Fukuda:2019ewf,Arvanitaki:2014wva}. We also considered the possibility of parametric resonance in the early universe and found that it is never important along the thermal-realignment DM line. We refer the reader to the End Matter for more details on these checks.

\paragraph{\textbf{Discussion and Conclusion.---}}
We have shown that an ultralight scalar DM with a sufficiently large Yukawa coupling to neutrinos and a sufficiently large amplitude at temperatures above the ballpark $T_*\sim 10\MeV-100\GeV$ would be thermally damped significantly by cosmic neutrinos. Thermal damping ameliorates late-time limits on such scalars, causes the scalar to approach a dynamical attractor leading to a predictable scalar abundance at the present epoch, and motivates a target line in the neutrino-scalar coupling vs scalar mass parameter space that hits the correct scalar DM abundance. These conclusions could, in principle, be generalized to thermal dampers other than the active neutrinos \footnote{These include but not limited to those considered in our companion paper \cite{LongPaper}.}. The attractor behavior should occur for any scalar $\phi$ with a strong amplitude $\Phi$-dependent thermal damping $\Upsilon(\Phi,T)$ such that (1) $\Phi(T)/\Phi_\Upsilon(T)$ decreases (increases) when $\Phi\gtrsim \Phi_\Upsilon$ ($\Phi\lesssim \Phi_{\Upsilon}$) and (2) the ratio $\Phi_\Upsilon(T)/[\Phi(T)|_{\Upsilon=0}]$ has a $\vee$ shape that dips below unity.

The particle physics ingredients considered here overlap with those of many other scenarios often considered in the literature. These include but are not limited to the so-called Mass Varying Neutrinos (MaVaNs) scenarios \cite{Fardon:2003eh,Kaplan:2004dq,Sakstein:2019fmf,Brookfield:2005bz,Kamionkowski:2024axz} and neutrino self-interactions (NSIs) scenarios which often include light mediators that might be dynamical in the early universe~\cite{Ge:2018uhz,Smirnov:2019cae,Babu:2019iml,Venzor:2020ova,Dutta:2022fdt,Berryman:2022hds,Kaplan:2024ydw,Green:2021gdc,Craig:2024tky,Esteban:2021ozz,Wang:2025qap}. 
The thermal damping of scalars by neutrinos we derived here may also be important in some parameter space of these models.

\paragraph*{\textbf{Acknowledgments.---}}
The authors would like to thank Xuheng Luo for collaboration in the early part of the project. The authors would also like to thank Anson Hook and David E. Kaplan for useful discussion. AB is supported by the National Science Foundation
under grant number PHY2210361 and the Maryland
Center for Fundamental Physics. 
E.H.T. is supported by NSF Grant PHY-2310429, Simons Investigator Award No. 824870, the Gordon and Betty Moore Foundation Grant GBMF7946, and the U.S. Department of Energy (DOE), Office of Science, National Quantum Information Science Research Centers, Superconducting Quantum Materials and Systems Center (SQMS) under contract No. DEAC02-07CH11359.

\section*{End Matter}

\subsection*{Various Checks}
\subsubsection{Thermalization of $\phi$ quanta}
If particle excitations of $\phi$ thermalizes with the neutrinos before BBN, it would add to the $\Delta N_{\rm eff}$ and spoil BBN. Neutrinos may emit $\phi$ quanta through weak-mediated bremsstrahlung $\nu\nu\rightarrow \nu\nu \phi$, with the rate $\Gamma_{\rm brem}\sim g^2G_F^2T^5$, for $T\lesssim 100\GeV$, which is potentially efficient at early times, at $T_{\rm brem}\sim 70\GeV(g/10^{-7})^{-2/3}$. In the regime $g\lesssim 10^{-7}$, in which our parameters of interest lie entirely, this estimate fails as it predicts $T_{\rm brem}\gtrsim 100\GeV$. At $T\gtrsim 100\GeV$, the bremsstrahlung rate becomes $\Gamma_{\rm brem}\propto T$, and there is no danger of $\phi$ thermalization then, as $\Gamma_{\rm brem}\propto T$ is a slower scaling than Hubble, $H\propto T^2$.

\subsubsection{Non-adiabatic neutrino production}
If $g\Phi>m_\nu$ then $M_\nu$ crosses zero every half-oscillation of $\phi$. Whenever that happens, the neutrinos' effective mass $M_\nu$ changes non-adiabatically, $|\dot{M}_\nu|/M_\nu^2\gtrsim 1$, and $\rho_\nu^{\rm NA}\sim |\dot{M}_\nu|^{2}$ worth of energy density in neutrinos is produced \cite{Kofman:1997yn,Tanin:2017bzm}. This leads to an effective friction $\Upsilon_{\rm NA}\sim M_\phi \rho_\nu^{\rm NA}/\rho_\phi\sim g^2 M_\phi$. To simplify our assumptions we avoid the regime where non-adiabatic particle production is significant. We require that $g\Phi> m_\nu$ and $g^2M_\phi\gtrsim H$ are never satisfied simultaneously. This amounts to requiring the latter at the latest time at which the former is satisfied, which boils down to $g\lesssim 9\times 10^{-9}\text{min}\left[1,(m_\phi/2\times 10^{-5}\eV)^{-2/13}\right]$.

\subsubsection{Parametric resonance due to loop-induced quartic coupling}
Since the first resonance band only populates non-relativistic $\phi$ modes which would quickly rejoin the condensate, we focus on the second band, which is centered around the momentum $k_{\rm res}\sim M_\phi$ with a width $\delta k_{\rm res}\sim M_\phi \sqrt{q}$ and Floquet exponent $\Gamma_{\rm Floquet}^{(2)}\sim M_\phi q^2$, where $q\sim (g^4/16\pi^2)\Phi^2/M_\phi^2$. Due to Hubble expansion, a given mode only stays in the resonance band for $\delta t_{\rm res}\sim H^{-1}\delta k_{\rm res}/k_{\rm res}\sim H^{-1}\sqrt{q}$. We therefore require $\Gamma_{\rm Floquet}^{(2)}\delta t_{\rm res}\lesssim 1$, which amounts to $g^4/16\pi^2\lesssim [M_\phi^4 H/\Phi^5]^{2/5}$, which is most important at the temperature $T\sim m_\phi/g$ when $M_\phi\sim m_\phi$, whereupon the condition reduces to $g\lesssim 1\times 10^{-8}(m_\phi/10^{-4}\eV)^{7/9}$, after setting $\Phi=\Phi_{\rm CDM}$.

\bibliography{references}

\begin{thebibliography}{87}%
\makeatletter
\providecommand \@ifxundefined [1]{%
 \@ifx{#1\undefined}
}%
\providecommand \@ifnum [1]{%
 \ifnum #1\expandafter \@firstoftwo
 \else \expandafter \@secondoftwo
 \fi
}%
\providecommand \@ifx [1]{%
 \ifx #1\expandafter \@firstoftwo
 \else \expandafter \@secondoftwo
 \fi
}%
\providecommand \natexlab [1]{#1}%
\providecommand \enquote  [1]{``#1''}%
\providecommand \bibnamefont  [1]{#1}%
\providecommand \bibfnamefont [1]{#1}%
\providecommand \citenamefont [1]{#1}%
\providecommand \href@noop [0]{\@secondoftwo}%
\providecommand \href [0]{\begingroup \@sanitize@url \@href}%
\providecommand \@href[1]{\@@startlink{#1}\@@href}%
\providecommand \@@href[1]{\endgroup#1\@@endlink}%
\providecommand \@sanitize@url [0]{\catcode `\\12\catcode `\$12\catcode `\&12\catcode `\#12\catcode `\^12\catcode `\_12\catcode `\%12\relax}%
\providecommand \@@startlink[1]{}%
\providecommand \@@endlink[0]{}%
\providecommand \url  [0]{\begingroup\@sanitize@url \@url }%
\providecommand \@url [1]{\endgroup\@href {#1}{\urlprefix }}%
\providecommand \urlprefix  [0]{URL }%
\providecommand \Eprint [0]{\href }%
\providecommand \doibase [0]{http://dx.doi.org/}%
\providecommand \selectlanguage [0]{\@gobble}%
\providecommand \bibinfo  [0]{\@secondoftwo}%
\providecommand \bibfield  [0]{\@secondoftwo}%
\providecommand \translation [1]{[#1]}%
\providecommand \BibitemOpen [0]{}%
\providecommand \bibitemStop [0]{}%
\providecommand \bibitemNoStop [0]{.\EOS\space}%
\providecommand \EOS [0]{\spacefactor3000\relax}%
\providecommand \BibitemShut  [1]{\csname bibitem#1\endcsname}%
\let\auto@bib@innerbib\@empty
\bibitem [{\citenamefont {Essig}\ \emph {et~al.}(2013)\citenamefont {Essig} \emph {et~al.}}]{Essig:2013lka}%
  \BibitemOpen
  \bibfield  {author} {\bibinfo {author} {\bibfnamefont {R.}~\bibnamefont {Essig}} \emph {et~al.},\ }in\ \href@noop {} {\emph {\bibinfo {booktitle} {{Snowmass 2013}: {Snowmass on the Mississippi}}}}\ (\bibinfo {year} {2013})\ \Eprint {http://arxiv.org/abs/1311.0029} {arXiv:1311.0029 [hep-ph]} \BibitemShut {NoStop}%
\bibitem [{\citenamefont {Ferreira}(2021)}]{Ferreira:2020fam}%
  \BibitemOpen
  \bibfield  {author} {\bibinfo {author} {\bibfnamefont {E.~G.~M.}\ \bibnamefont {Ferreira}},\ }\href {\doibase 10.1007/s00159-021-00135-6} {\bibfield  {journal} {\bibinfo  {journal} {Astron. Astrophys. Rev.}\ }\textbf {\bibinfo {volume} {29}},\ \bibinfo {pages} {7} (\bibinfo {year} {2021})},\ \Eprint {http://arxiv.org/abs/2005.03254} {arXiv:2005.03254 [astro-ph.CO]} \BibitemShut {NoStop}%
\bibitem [{\citenamefont {Hui}\ \emph {et~al.}(2017)\citenamefont {Hui}, \citenamefont {Ostriker}, \citenamefont {Tremaine},\ and\ \citenamefont {Witten}}]{Hui:2016ltb}%
  \BibitemOpen
  \bibfield  {author} {\bibinfo {author} {\bibfnamefont {L.}~\bibnamefont {Hui}}, \bibinfo {author} {\bibfnamefont {J.~P.}\ \bibnamefont {Ostriker}}, \bibinfo {author} {\bibfnamefont {S.}~\bibnamefont {Tremaine}}, \ and\ \bibinfo {author} {\bibfnamefont {E.}~\bibnamefont {Witten}},\ }\href {\doibase 10.1103/PhysRevD.95.043541} {\bibfield  {journal} {\bibinfo  {journal} {Phys. Rev. D}\ }\textbf {\bibinfo {volume} {95}},\ \bibinfo {pages} {043541} (\bibinfo {year} {2017})},\ \Eprint {http://arxiv.org/abs/1610.08297} {arXiv:1610.08297 [astro-ph.CO]} \BibitemShut {NoStop}%
\bibitem [{\citenamefont {Berlin}(2016)}]{Berlin:2016woy}%
  \BibitemOpen
  \bibfield  {author} {\bibinfo {author} {\bibfnamefont {A.}~\bibnamefont {Berlin}},\ }\href {\doibase 10.1103/PhysRevLett.117.231801} {\bibfield  {journal} {\bibinfo  {journal} {Phys. Rev. Lett.}\ }\textbf {\bibinfo {volume} {117}},\ \bibinfo {pages} {231801} (\bibinfo {year} {2016})},\ \Eprint {http://arxiv.org/abs/1608.01307} {arXiv:1608.01307 [hep-ph]} \BibitemShut {NoStop}%
\bibitem [{\citenamefont {Krnjaic}\ \emph {et~al.}(2018)\citenamefont {Krnjaic}, \citenamefont {Machado},\ and\ \citenamefont {Necib}}]{Krnjaic:2017zlz}%
  \BibitemOpen
  \bibfield  {author} {\bibinfo {author} {\bibfnamefont {G.}~\bibnamefont {Krnjaic}}, \bibinfo {author} {\bibfnamefont {P.~A.~N.}\ \bibnamefont {Machado}}, \ and\ \bibinfo {author} {\bibfnamefont {L.}~\bibnamefont {Necib}},\ }\href {\doibase 10.1103/PhysRevD.97.075017} {\bibfield  {journal} {\bibinfo  {journal} {Phys. Rev. D}\ }\textbf {\bibinfo {volume} {97}},\ \bibinfo {pages} {075017} (\bibinfo {year} {2018})},\ \Eprint {http://arxiv.org/abs/1705.06740} {arXiv:1705.06740 [hep-ph]} \BibitemShut {NoStop}%
\bibitem [{\citenamefont {Brdar}\ \emph {et~al.}(2018)\citenamefont {Brdar}, \citenamefont {Kopp}, \citenamefont {Liu}, \citenamefont {Prass},\ and\ \citenamefont {Wang}}]{Brdar:2017kbt}%
  \BibitemOpen
  \bibfield  {author} {\bibinfo {author} {\bibfnamefont {V.}~\bibnamefont {Brdar}}, \bibinfo {author} {\bibfnamefont {J.}~\bibnamefont {Kopp}}, \bibinfo {author} {\bibfnamefont {J.}~\bibnamefont {Liu}}, \bibinfo {author} {\bibfnamefont {P.}~\bibnamefont {Prass}}, \ and\ \bibinfo {author} {\bibfnamefont {X.-P.}\ \bibnamefont {Wang}},\ }\href {\doibase 10.1103/PhysRevD.97.043001} {\bibfield  {journal} {\bibinfo  {journal} {Phys. Rev. D}\ }\textbf {\bibinfo {volume} {97}},\ \bibinfo {pages} {043001} (\bibinfo {year} {2018})},\ \Eprint {http://arxiv.org/abs/1705.09455} {arXiv:1705.09455 [hep-ph]} \BibitemShut {NoStop}%
\bibitem [{\citenamefont {Liao}\ \emph {et~al.}(2018)\citenamefont {Liao}, \citenamefont {Marfatia},\ and\ \citenamefont {Whisnant}}]{Liao:2018byh}%
  \BibitemOpen
  \bibfield  {author} {\bibinfo {author} {\bibfnamefont {J.}~\bibnamefont {Liao}}, \bibinfo {author} {\bibfnamefont {D.}~\bibnamefont {Marfatia}}, \ and\ \bibinfo {author} {\bibfnamefont {K.}~\bibnamefont {Whisnant}},\ }\href {\doibase 10.1007/JHEP04(2018)136} {\bibfield  {journal} {\bibinfo  {journal} {JHEP}\ }\textbf {\bibinfo {volume} {04}},\ \bibinfo {pages} {136} (\bibinfo {year} {2018})},\ \Eprint {http://arxiv.org/abs/1803.01773} {arXiv:1803.01773 [hep-ph]} \BibitemShut {NoStop}%
\bibitem [{\citenamefont {Capozzi}\ \emph {et~al.}(2018)\citenamefont {Capozzi}, \citenamefont {Shoemaker},\ and\ \citenamefont {Vecchi}}]{Capozzi:2018bps}%
  \BibitemOpen
  \bibfield  {author} {\bibinfo {author} {\bibfnamefont {F.}~\bibnamefont {Capozzi}}, \bibinfo {author} {\bibfnamefont {I.~M.}\ \bibnamefont {Shoemaker}}, \ and\ \bibinfo {author} {\bibfnamefont {L.}~\bibnamefont {Vecchi}},\ }\href {\doibase 10.1088/1475-7516/2018/07/004} {\bibfield  {journal} {\bibinfo  {journal} {JCAP}\ }\textbf {\bibinfo {volume} {07}},\ \bibinfo {pages} {004} (\bibinfo {year} {2018})},\ \Eprint {http://arxiv.org/abs/1804.05117} {arXiv:1804.05117 [hep-ph]} \BibitemShut {NoStop}%
\bibitem [{\citenamefont {Huang}\ and\ \citenamefont {Nath}(2018)}]{Huang:2018cwo}%
  \BibitemOpen
  \bibfield  {author} {\bibinfo {author} {\bibfnamefont {G.-Y.}\ \bibnamefont {Huang}}\ and\ \bibinfo {author} {\bibfnamefont {N.}~\bibnamefont {Nath}},\ }\href {\doibase 10.1140/epjc/s10052-018-6391-y} {\bibfield  {journal} {\bibinfo  {journal} {Eur. Phys. J. C}\ }\textbf {\bibinfo {volume} {78}},\ \bibinfo {pages} {922} (\bibinfo {year} {2018})},\ \Eprint {http://arxiv.org/abs/1809.01111} {arXiv:1809.01111 [hep-ph]} \BibitemShut {NoStop}%
\bibitem [{\citenamefont {Cline}(2020)}]{Cline:2019seo}%
  \BibitemOpen
  \bibfield  {author} {\bibinfo {author} {\bibfnamefont {J.~M.}\ \bibnamefont {Cline}},\ }\href {\doibase 10.1016/j.physletb.2019.135182} {\bibfield  {journal} {\bibinfo  {journal} {Phys. Lett. B}\ }\textbf {\bibinfo {volume} {802}},\ \bibinfo {pages} {135182} (\bibinfo {year} {2020})},\ \Eprint {http://arxiv.org/abs/1908.02278} {arXiv:1908.02278 [hep-ph]} \BibitemShut {NoStop}%
\bibitem [{\citenamefont {Dev}\ \emph {et~al.}(2021)\citenamefont {Dev}, \citenamefont {Machado},\ and\ \citenamefont {Mart{\'\i}nez-Mirav{\'e}}}]{Dev:2020kgz}%
  \BibitemOpen
  \bibfield  {author} {\bibinfo {author} {\bibfnamefont {A.}~\bibnamefont {Dev}}, \bibinfo {author} {\bibfnamefont {P.~A.~N.}\ \bibnamefont {Machado}}, \ and\ \bibinfo {author} {\bibfnamefont {P.}~\bibnamefont {Mart{\'\i}nez-Mirav{\'e}}},\ }\href {\doibase 10.1007/JHEP01(2021)094} {\bibfield  {journal} {\bibinfo  {journal} {JHEP}\ }\textbf {\bibinfo {volume} {01}},\ \bibinfo {pages} {094} (\bibinfo {year} {2021})},\ \Eprint {http://arxiv.org/abs/2007.03590} {arXiv:2007.03590 [hep-ph]} \BibitemShut {NoStop}%
\bibitem [{\citenamefont {Huang}\ and\ \citenamefont {Nath}(2022)}]{Huang:2021kam}%
  \BibitemOpen
  \bibfield  {author} {\bibinfo {author} {\bibfnamefont {G.-y.}\ \bibnamefont {Huang}}\ and\ \bibinfo {author} {\bibfnamefont {N.}~\bibnamefont {Nath}},\ }\href {\doibase 10.1088/1475-7516/2022/05/034} {\bibfield  {journal} {\bibinfo  {journal} {JCAP}\ }\textbf {\bibinfo {volume} {05}},\ \bibinfo {pages} {034} (\bibinfo {year} {2022})},\ \Eprint {http://arxiv.org/abs/2111.08732} {arXiv:2111.08732 [hep-ph]} \BibitemShut {NoStop}%
\bibitem [{\citenamefont {Losada}\ \emph {et~al.}(2022)\citenamefont {Losada}, \citenamefont {Nir}, \citenamefont {Perez},\ and\ \citenamefont {Shpilman}}]{Losada:2021bxx}%
  \BibitemOpen
  \bibfield  {author} {\bibinfo {author} {\bibfnamefont {M.}~\bibnamefont {Losada}}, \bibinfo {author} {\bibfnamefont {Y.}~\bibnamefont {Nir}}, \bibinfo {author} {\bibfnamefont {G.}~\bibnamefont {Perez}}, \ and\ \bibinfo {author} {\bibfnamefont {Y.}~\bibnamefont {Shpilman}},\ }\href {\doibase 10.1007/JHEP04(2022)030} {\bibfield  {journal} {\bibinfo  {journal} {JHEP}\ }\textbf {\bibinfo {volume} {04}},\ \bibinfo {pages} {030} (\bibinfo {year} {2022})},\ \Eprint {http://arxiv.org/abs/2107.10865} {arXiv:2107.10865 [hep-ph]} \BibitemShut {NoStop}%
\bibitem [{\citenamefont {Chun}(2021)}]{Chun:2021ief}%
  \BibitemOpen
  \bibfield  {author} {\bibinfo {author} {\bibfnamefont {E.~J.}\ \bibnamefont {Chun}},\ }\href@noop {} {\  (\bibinfo {year} {2021})},\ \Eprint {http://arxiv.org/abs/2112.05057} {arXiv:2112.05057 [hep-ph]} \BibitemShut {NoStop}%
\bibitem [{\citenamefont {Dev}\ \emph {et~al.}(2023)\citenamefont {Dev}, \citenamefont {Krnjaic}, \citenamefont {Machado},\ and\ \citenamefont {Ramani}}]{Dev:2022bae}%
  \BibitemOpen
  \bibfield  {author} {\bibinfo {author} {\bibfnamefont {A.}~\bibnamefont {Dev}}, \bibinfo {author} {\bibfnamefont {G.}~\bibnamefont {Krnjaic}}, \bibinfo {author} {\bibfnamefont {P.}~\bibnamefont {Machado}}, \ and\ \bibinfo {author} {\bibfnamefont {H.}~\bibnamefont {Ramani}},\ }\href {\doibase 10.1103/PhysRevD.107.035006} {\bibfield  {journal} {\bibinfo  {journal} {Phys. Rev. D}\ }\textbf {\bibinfo {volume} {107}},\ \bibinfo {pages} {035006} (\bibinfo {year} {2023})},\ \Eprint {http://arxiv.org/abs/2205.06821} {arXiv:2205.06821 [hep-ph]} \BibitemShut {NoStop}%
\bibitem [{\citenamefont {Huang}\ \emph {et~al.}(2022)\citenamefont {Huang}, \citenamefont {Lindner}, \citenamefont {Mart{\'\i}nez-Mirav{\'e}},\ and\ \citenamefont {Sen}}]{Huang:2022wmz}%
  \BibitemOpen
  \bibfield  {author} {\bibinfo {author} {\bibfnamefont {G.-y.}\ \bibnamefont {Huang}}, \bibinfo {author} {\bibfnamefont {M.}~\bibnamefont {Lindner}}, \bibinfo {author} {\bibfnamefont {P.}~\bibnamefont {Mart{\'\i}nez-Mirav{\'e}}}, \ and\ \bibinfo {author} {\bibfnamefont {M.}~\bibnamefont {Sen}},\ }\href {\doibase 10.1103/PhysRevD.106.033004} {\bibfield  {journal} {\bibinfo  {journal} {Phys. Rev. D}\ }\textbf {\bibinfo {volume} {106}},\ \bibinfo {pages} {033004} (\bibinfo {year} {2022})},\ \Eprint {http://arxiv.org/abs/2205.08431} {arXiv:2205.08431 [hep-ph]} \BibitemShut {NoStop}%
\bibitem [{\citenamefont {Losada}\ \emph {et~al.}(2023)\citenamefont {Losada}, \citenamefont {Nir}, \citenamefont {Perez}, \citenamefont {Savoray},\ and\ \citenamefont {Shpilman}}]{Losada:2022uvr}%
  \BibitemOpen
  \bibfield  {author} {\bibinfo {author} {\bibfnamefont {M.}~\bibnamefont {Losada}}, \bibinfo {author} {\bibfnamefont {Y.}~\bibnamefont {Nir}}, \bibinfo {author} {\bibfnamefont {G.}~\bibnamefont {Perez}}, \bibinfo {author} {\bibfnamefont {I.}~\bibnamefont {Savoray}}, \ and\ \bibinfo {author} {\bibfnamefont {Y.}~\bibnamefont {Shpilman}},\ }\href {\doibase 10.1007/JHEP03(2023)032} {\bibfield  {journal} {\bibinfo  {journal} {JHEP}\ }\textbf {\bibinfo {volume} {03}},\ \bibinfo {pages} {032} (\bibinfo {year} {2023})},\ \Eprint {http://arxiv.org/abs/2205.09769} {arXiv:2205.09769 [hep-ph]} \BibitemShut {NoStop}%
\bibitem [{\citenamefont {Plestid}\ and\ \citenamefont {Tevosyan}(2025)}]{Plestid:2024kyy}%
  \BibitemOpen
  \bibfield  {author} {\bibinfo {author} {\bibfnamefont {R.}~\bibnamefont {Plestid}}\ and\ \bibinfo {author} {\bibfnamefont {S.}~\bibnamefont {Tevosyan}},\ }\href {\doibase 10.1007/JHEP07(2025)012} {\bibfield  {journal} {\bibinfo  {journal} {JHEP}\ }\textbf {\bibinfo {volume} {07}},\ \bibinfo {pages} {012} (\bibinfo {year} {2025})},\ \Eprint {http://arxiv.org/abs/2409.17396} {arXiv:2409.17396 [hep-ph]} \BibitemShut {NoStop}%
\bibitem [{\citenamefont {Gherghetta}\ and\ \citenamefont {Shkerin}(2023)}]{Gherghetta:2023myo}%
  \BibitemOpen
  \bibfield  {author} {\bibinfo {author} {\bibfnamefont {T.}~\bibnamefont {Gherghetta}}\ and\ \bibinfo {author} {\bibfnamefont {A.}~\bibnamefont {Shkerin}},\ }\href {\doibase 10.1103/PhysRevD.108.095009} {\bibfield  {journal} {\bibinfo  {journal} {Phys. Rev. D}\ }\textbf {\bibinfo {volume} {108}},\ \bibinfo {pages} {095009} (\bibinfo {year} {2023})},\ \Eprint {http://arxiv.org/abs/2305.06441} {arXiv:2305.06441 [hep-ph]} \BibitemShut {NoStop}%
\bibitem [{\citenamefont {Brzeminski}\ \emph {et~al.}(2023)\citenamefont {Brzeminski}, \citenamefont {Das}, \citenamefont {Hook},\ and\ \citenamefont {Ristow}}]{Brzeminski:2022rkf}%
  \BibitemOpen
  \bibfield  {author} {\bibinfo {author} {\bibfnamefont {D.}~\bibnamefont {Brzeminski}}, \bibinfo {author} {\bibfnamefont {S.}~\bibnamefont {Das}}, \bibinfo {author} {\bibfnamefont {A.}~\bibnamefont {Hook}}, \ and\ \bibinfo {author} {\bibfnamefont {C.}~\bibnamefont {Ristow}},\ }\href {\doibase 10.1007/JHEP08(2023)181} {\bibfield  {journal} {\bibinfo  {journal} {JHEP}\ }\textbf {\bibinfo {volume} {08}},\ \bibinfo {pages} {181} (\bibinfo {year} {2023})},\ \Eprint {http://arxiv.org/abs/2212.05073} {arXiv:2212.05073 [hep-ph]} \BibitemShut {NoStop}%
\bibitem [{\citenamefont {Tanin}\ and\ \citenamefont {Stewart}(2017)}]{Tanin:2017bzm}%
  \BibitemOpen
  \bibfield  {author} {\bibinfo {author} {\bibfnamefont {E.~H.}\ \bibnamefont {Tanin}}\ and\ \bibinfo {author} {\bibfnamefont {E.~D.}\ \bibnamefont {Stewart}},\ }\href {\doibase 10.1088/1475-7516/2017/11/019} {\bibfield  {journal} {\bibinfo  {journal} {JCAP}\ }\textbf {\bibinfo {volume} {11}},\ \bibinfo {pages} {019} (\bibinfo {year} {2017})},\ \Eprint {http://arxiv.org/abs/1708.04865} {arXiv:1708.04865 [hep-ph]} \BibitemShut {NoStop}%
\bibitem [{\citenamefont {Mukaida}\ and\ \citenamefont {Nakayama}(2013{\natexlab{a}})}]{Mukaida:2012qn}%
  \BibitemOpen
  \bibfield  {author} {\bibinfo {author} {\bibfnamefont {K.}~\bibnamefont {Mukaida}}\ and\ \bibinfo {author} {\bibfnamefont {K.}~\bibnamefont {Nakayama}},\ }\href {\doibase 10.1088/1475-7516/2013/01/017} {\bibfield  {journal} {\bibinfo  {journal} {JCAP}\ }\textbf {\bibinfo {volume} {01}},\ \bibinfo {pages} {017} (\bibinfo {year} {2013}{\natexlab{a}})},\ \Eprint {http://arxiv.org/abs/1208.3399} {arXiv:1208.3399 [hep-ph]} \BibitemShut {NoStop}%
\bibitem [{\citenamefont {Mukaida}\ and\ \citenamefont {Nakayama}(2013{\natexlab{b}})}]{Mukaida:2012bz}%
  \BibitemOpen
  \bibfield  {author} {\bibinfo {author} {\bibfnamefont {K.}~\bibnamefont {Mukaida}}\ and\ \bibinfo {author} {\bibfnamefont {K.}~\bibnamefont {Nakayama}},\ }\href {\doibase 10.1088/1475-7516/2013/03/002} {\bibfield  {journal} {\bibinfo  {journal} {JCAP}\ }\textbf {\bibinfo {volume} {03}},\ \bibinfo {pages} {002} (\bibinfo {year} {2013}{\natexlab{b}})},\ \Eprint {http://arxiv.org/abs/1212.4985} {arXiv:1212.4985 [hep-ph]} \BibitemShut {NoStop}%
\bibitem [{\citenamefont {Yokoyama}(2004)}]{Yokoyama:2004pf}%
  \BibitemOpen
  \bibfield  {author} {\bibinfo {author} {\bibfnamefont {J.}~\bibnamefont {Yokoyama}},\ }\href {\doibase 10.1103/PhysRevD.70.103511} {\bibfield  {journal} {\bibinfo  {journal} {Phys. Rev. D}\ }\textbf {\bibinfo {volume} {70}},\ \bibinfo {pages} {103511} (\bibinfo {year} {2004})},\ \Eprint {http://arxiv.org/abs/hep-ph/0406072} {arXiv:hep-ph/0406072} \BibitemShut {NoStop}%
\bibitem [{\citenamefont {Yokoyama}\ and\ \citenamefont {Linde}(1999)}]{Yokoyama:1998ju}%
  \BibitemOpen
  \bibfield  {author} {\bibinfo {author} {\bibfnamefont {J.}~\bibnamefont {Yokoyama}}\ and\ \bibinfo {author} {\bibfnamefont {A.~D.}\ \bibnamefont {Linde}},\ }\href {\doibase 10.1103/PhysRevD.60.083509} {\bibfield  {journal} {\bibinfo  {journal} {Phys. Rev. D}\ }\textbf {\bibinfo {volume} {60}},\ \bibinfo {pages} {083509} (\bibinfo {year} {1999})},\ \Eprint {http://arxiv.org/abs/hep-ph/9809409} {arXiv:hep-ph/9809409} \BibitemShut {NoStop}%
\bibitem [{\citenamefont {Bastero-Gil}\ \emph {et~al.}(2011)\citenamefont {Bastero-Gil}, \citenamefont {Berera},\ and\ \citenamefont {Ramos}}]{Bastero-Gil:2010dgy}%
  \BibitemOpen
  \bibfield  {author} {\bibinfo {author} {\bibfnamefont {M.}~\bibnamefont {Bastero-Gil}}, \bibinfo {author} {\bibfnamefont {A.}~\bibnamefont {Berera}}, \ and\ \bibinfo {author} {\bibfnamefont {R.~O.}\ \bibnamefont {Ramos}},\ }\href {\doibase 10.1088/1475-7516/2011/09/033} {\bibfield  {journal} {\bibinfo  {journal} {JCAP}\ }\textbf {\bibinfo {volume} {09}},\ \bibinfo {pages} {033} (\bibinfo {year} {2011})},\ \Eprint {http://arxiv.org/abs/1008.1929} {arXiv:1008.1929 [hep-ph]} \BibitemShut {NoStop}%
\bibitem [{\citenamefont {B{\"o}deker}\ and\ \citenamefont {Nienaber}(2022)}]{Bodeker:2022ihg}%
  \BibitemOpen
  \bibfield  {author} {\bibinfo {author} {\bibfnamefont {D.}~\bibnamefont {B{\"o}deker}}\ and\ \bibinfo {author} {\bibfnamefont {J.}~\bibnamefont {Nienaber}},\ }\href {\doibase 10.1103/PhysRevD.106.056016} {\bibfield  {journal} {\bibinfo  {journal} {Phys. Rev. D}\ }\textbf {\bibinfo {volume} {106}},\ \bibinfo {pages} {056016} (\bibinfo {year} {2022})},\ \Eprint {http://arxiv.org/abs/2205.14166} {arXiv:2205.14166 [hep-ph]} \BibitemShut {NoStop}%
\bibitem [{\citenamefont {Hosoya}\ and\ \citenamefont {Sakagami}(1984)}]{Hosoya:1983ke}%
  \BibitemOpen
  \bibfield  {author} {\bibinfo {author} {\bibfnamefont {A.}~\bibnamefont {Hosoya}}\ and\ \bibinfo {author} {\bibfnamefont {M.-a.}\ \bibnamefont {Sakagami}},\ }\href {\doibase 10.1103/PhysRevD.29.2228} {\bibfield  {journal} {\bibinfo  {journal} {Phys. Rev. D}\ }\textbf {\bibinfo {volume} {29}},\ \bibinfo {pages} {2228} (\bibinfo {year} {1984})}\BibitemShut {NoStop}%
\bibitem [{Note1()}]{Note1}%
  \BibitemOpen
  \bibinfo {note} {We discuss thermal damping more generally in a companion paper~\cite {LongPaper}.}\BibitemShut {Stop}%
\bibitem [{\citenamefont {Affleck}\ and\ \citenamefont {Dine}(1985)}]{Affleck:1984fy}%
  \BibitemOpen
  \bibfield  {author} {\bibinfo {author} {\bibfnamefont {I.}~\bibnamefont {Affleck}}\ and\ \bibinfo {author} {\bibfnamefont {M.}~\bibnamefont {Dine}},\ }\href {\doibase 10.1016/0550-3213(85)90021-5} {\bibfield  {journal} {\bibinfo  {journal} {Nucl. Phys. B}\ }\textbf {\bibinfo {volume} {249}},\ \bibinfo {pages} {361} (\bibinfo {year} {1985})}\BibitemShut {NoStop}%
\bibitem [{\citenamefont {Stewart}\ \emph {et~al.}(1996)\citenamefont {Stewart}, \citenamefont {Kawasaki},\ and\ \citenamefont {Yanagida}}]{Stewart:1996ai}%
  \BibitemOpen
  \bibfield  {author} {\bibinfo {author} {\bibfnamefont {E.~D.}\ \bibnamefont {Stewart}}, \bibinfo {author} {\bibfnamefont {M.}~\bibnamefont {Kawasaki}}, \ and\ \bibinfo {author} {\bibfnamefont {T.}~\bibnamefont {Yanagida}},\ }\href {\doibase 10.1103/PhysRevD.54.6032} {\bibfield  {journal} {\bibinfo  {journal} {Phys. Rev. D}\ }\textbf {\bibinfo {volume} {54}},\ \bibinfo {pages} {6032} (\bibinfo {year} {1996})},\ \Eprint {http://arxiv.org/abs/hep-ph/9603324} {arXiv:hep-ph/9603324} \BibitemShut {NoStop}%
\bibitem [{\citenamefont {Dine}\ \emph {et~al.}(1996)\citenamefont {Dine}, \citenamefont {Randall},\ and\ \citenamefont {Thomas}}]{Dine:1995kz}%
  \BibitemOpen
  \bibfield  {author} {\bibinfo {author} {\bibfnamefont {M.}~\bibnamefont {Dine}}, \bibinfo {author} {\bibfnamefont {L.}~\bibnamefont {Randall}}, \ and\ \bibinfo {author} {\bibfnamefont {S.~D.}\ \bibnamefont {Thomas}},\ }\href {\doibase 10.1016/0550-3213(95)00538-2} {\bibfield  {journal} {\bibinfo  {journal} {Nucl. Phys. B}\ }\textbf {\bibinfo {volume} {458}},\ \bibinfo {pages} {291} (\bibinfo {year} {1996})},\ \Eprint {http://arxiv.org/abs/hep-ph/9507453} {arXiv:hep-ph/9507453} \BibitemShut {NoStop}%
\bibitem [{\citenamefont {Chang}\ \emph {et~al.}(2025)\citenamefont {Chang}, \citenamefont {Shin},\ and\ \citenamefont {Unwin}}]{Chang:2025eef}%
  \BibitemOpen
  \bibfield  {author} {\bibinfo {author} {\bibfnamefont {J.~H.}\ \bibnamefont {Chang}}, \bibinfo {author} {\bibfnamefont {C.~S.}\ \bibnamefont {Shin}}, \ and\ \bibinfo {author} {\bibfnamefont {J.}~\bibnamefont {Unwin}},\ }\href@noop {} {\  (\bibinfo {year} {2025})},\ \Eprint {http://arxiv.org/abs/2506.03269} {arXiv:2506.03269 [hep-ph]} \BibitemShut {NoStop}%
\bibitem [{\citenamefont {Chang}\ \emph {et~al.}(2024)\citenamefont {Chang}, \citenamefont {Jeong}, \citenamefont {Lee},\ and\ \citenamefont {Shin}}]{Chang:2024xjd}%
  \BibitemOpen
  \bibfield  {author} {\bibinfo {author} {\bibfnamefont {J.~H.}\ \bibnamefont {Chang}}, \bibinfo {author} {\bibfnamefont {K.~S.}\ \bibnamefont {Jeong}}, \bibinfo {author} {\bibfnamefont {C.~H.}\ \bibnamefont {Lee}}, \ and\ \bibinfo {author} {\bibfnamefont {C.~S.}\ \bibnamefont {Shin}},\ }\href {\doibase 10.1103/PhysRevD.110.055038} {\bibfield  {journal} {\bibinfo  {journal} {Phys. Rev. D}\ }\textbf {\bibinfo {volume} {110}},\ \bibinfo {pages} {055038} (\bibinfo {year} {2024})},\ \Eprint {http://arxiv.org/abs/2401.13734} {arXiv:2401.13734 [hep-ph]} \BibitemShut {NoStop}%
\bibitem [{\citenamefont {Chang}\ \emph {et~al.}(2022)\citenamefont {Chang}, \citenamefont {Olea-Romacho},\ and\ \citenamefont {Tanin}}]{Chang:2022psj}%
  \BibitemOpen
  \bibfield  {author} {\bibinfo {author} {\bibfnamefont {J.~H.}\ \bibnamefont {Chang}}, \bibinfo {author} {\bibfnamefont {M.~O.}\ \bibnamefont {Olea-Romacho}}, \ and\ \bibinfo {author} {\bibfnamefont {E.~H.}\ \bibnamefont {Tanin}},\ }\href {\doibase 10.1103/PhysRevD.106.113003} {\bibfield  {journal} {\bibinfo  {journal} {Phys. Rev. D}\ }\textbf {\bibinfo {volume} {106}},\ \bibinfo {pages} {113003} (\bibinfo {year} {2022})},\ \Eprint {http://arxiv.org/abs/2210.05680} {arXiv:2210.05680 [hep-ph]} \BibitemShut {NoStop}%
\bibitem [{\citenamefont {Banks}\ \emph {et~al.}(1994)\citenamefont {Banks}, \citenamefont {Kaplan},\ and\ \citenamefont {Nelson}}]{Banks:1993en}%
  \BibitemOpen
  \bibfield  {author} {\bibinfo {author} {\bibfnamefont {T.}~\bibnamefont {Banks}}, \bibinfo {author} {\bibfnamefont {D.~B.}\ \bibnamefont {Kaplan}}, \ and\ \bibinfo {author} {\bibfnamefont {A.~E.}\ \bibnamefont {Nelson}},\ }\href {\doibase 10.1103/PhysRevD.49.779} {\bibfield  {journal} {\bibinfo  {journal} {Phys. Rev. D}\ }\textbf {\bibinfo {volume} {49}},\ \bibinfo {pages} {779} (\bibinfo {year} {1994})},\ \Eprint {http://arxiv.org/abs/hep-ph/9308292} {arXiv:hep-ph/9308292} \BibitemShut {NoStop}%
\bibitem [{\citenamefont {Lyth}\ and\ \citenamefont {Stewart}(1996)}]{Lyth:1995ka}%
  \BibitemOpen
  \bibfield  {author} {\bibinfo {author} {\bibfnamefont {D.~H.}\ \bibnamefont {Lyth}}\ and\ \bibinfo {author} {\bibfnamefont {E.~D.}\ \bibnamefont {Stewart}},\ }\href {\doibase 10.1103/PhysRevD.53.1784} {\bibfield  {journal} {\bibinfo  {journal} {Phys. Rev. D}\ }\textbf {\bibinfo {volume} {53}},\ \bibinfo {pages} {1784} (\bibinfo {year} {1996})},\ \Eprint {http://arxiv.org/abs/hep-ph/9510204} {arXiv:hep-ph/9510204} \BibitemShut {NoStop}%
\bibitem [{\citenamefont {Batell}\ and\ \citenamefont {Ghalsasi}(2023)}]{Batell:2021ofv}%
  \BibitemOpen
  \bibfield  {author} {\bibinfo {author} {\bibfnamefont {B.}~\bibnamefont {Batell}}\ and\ \bibinfo {author} {\bibfnamefont {A.}~\bibnamefont {Ghalsasi}},\ }\href {\doibase 10.1103/PhysRevD.107.L091701} {\bibfield  {journal} {\bibinfo  {journal} {Phys. Rev. D}\ }\textbf {\bibinfo {volume} {107}},\ \bibinfo {pages} {L091701} (\bibinfo {year} {2023})},\ \Eprint {http://arxiv.org/abs/2109.04476} {arXiv:2109.04476 [hep-ph]} \BibitemShut {NoStop}%
\bibitem [{\citenamefont {Batell}\ \emph {et~al.}(2024)\citenamefont {Batell}, \citenamefont {Ghalsasi},\ and\ \citenamefont {Rai}}]{Batell:2022qvr}%
  \BibitemOpen
  \bibfield  {author} {\bibinfo {author} {\bibfnamefont {B.}~\bibnamefont {Batell}}, \bibinfo {author} {\bibfnamefont {A.}~\bibnamefont {Ghalsasi}}, \ and\ \bibinfo {author} {\bibfnamefont {M.}~\bibnamefont {Rai}},\ }\href {\doibase 10.1007/JHEP01(2024)038} {\bibfield  {journal} {\bibinfo  {journal} {JHEP}\ }\textbf {\bibinfo {volume} {01}},\ \bibinfo {pages} {038} (\bibinfo {year} {2024})},\ \Eprint {http://arxiv.org/abs/2211.09132} {arXiv:2211.09132 [hep-ph]} \BibitemShut {NoStop}%
\bibitem [{\citenamefont {Cyncynates}\ and\ \citenamefont {Simon}(2025)}]{Cyncynates:2024bxw}%
  \BibitemOpen
  \bibfield  {author} {\bibinfo {author} {\bibfnamefont {D.}~\bibnamefont {Cyncynates}}\ and\ \bibinfo {author} {\bibfnamefont {O.}~\bibnamefont {Simon}},\ }\href {\doibase 10.1103/8mc9-6cmr} {\bibfield  {journal} {\bibinfo  {journal} {Phys. Rev. D}\ }\textbf {\bibinfo {volume} {112}},\ \bibinfo {pages} {055002} (\bibinfo {year} {2025})},\ \Eprint {http://arxiv.org/abs/2408.16816} {arXiv:2408.16816 [hep-ph]} \BibitemShut {NoStop}%
\bibitem [{\citenamefont {Jungman}\ \emph {et~al.}(1996)\citenamefont {Jungman}, \citenamefont {Kamionkowski},\ and\ \citenamefont {Griest}}]{Jungman:1995df}%
  \BibitemOpen
  \bibfield  {author} {\bibinfo {author} {\bibfnamefont {G.}~\bibnamefont {Jungman}}, \bibinfo {author} {\bibfnamefont {M.}~\bibnamefont {Kamionkowski}}, \ and\ \bibinfo {author} {\bibfnamefont {K.}~\bibnamefont {Griest}},\ }\href {\doibase 10.1016/0370-1573(95)00058-5} {\bibfield  {journal} {\bibinfo  {journal} {Phys. Rept.}\ }\textbf {\bibinfo {volume} {267}},\ \bibinfo {pages} {195} (\bibinfo {year} {1996})},\ \Eprint {http://arxiv.org/abs/hep-ph/9506380} {arXiv:hep-ph/9506380} \BibitemShut {NoStop}%
\bibitem [{\citenamefont {Hall}\ \emph {et~al.}(2010)\citenamefont {Hall}, \citenamefont {Jedamzik}, \citenamefont {March-Russell},\ and\ \citenamefont {West}}]{Hall:2009bx}%
  \BibitemOpen
  \bibfield  {author} {\bibinfo {author} {\bibfnamefont {L.~J.}\ \bibnamefont {Hall}}, \bibinfo {author} {\bibfnamefont {K.}~\bibnamefont {Jedamzik}}, \bibinfo {author} {\bibfnamefont {J.}~\bibnamefont {March-Russell}}, \ and\ \bibinfo {author} {\bibfnamefont {S.~M.}\ \bibnamefont {West}},\ }\href {\doibase 10.1007/JHEP03(2010)080} {\bibfield  {journal} {\bibinfo  {journal} {JHEP}\ }\textbf {\bibinfo {volume} {03}},\ \bibinfo {pages} {080} (\bibinfo {year} {2010})},\ \Eprint {http://arxiv.org/abs/0911.1120} {arXiv:0911.1120 [hep-ph]} \BibitemShut {NoStop}%
\bibitem [{\citenamefont {McDonald}(2002)}]{McDonald:2001vt}%
  \BibitemOpen
  \bibfield  {author} {\bibinfo {author} {\bibfnamefont {J.}~\bibnamefont {McDonald}},\ }\href {\doibase 10.1103/PhysRevLett.88.091304} {\bibfield  {journal} {\bibinfo  {journal} {Phys. Rev. Lett.}\ }\textbf {\bibinfo {volume} {88}},\ \bibinfo {pages} {091304} (\bibinfo {year} {2002})},\ \Eprint {http://arxiv.org/abs/hep-ph/0106249} {arXiv:hep-ph/0106249} \BibitemShut {NoStop}%
\bibitem [{\citenamefont {{Bhatnagar}}\ \emph {et~al.}(1954)\citenamefont {{Bhatnagar}}, \citenamefont {{Gross}},\ and\ \citenamefont {{Krook}}}]{1954PhRv...94..511B}%
  \BibitemOpen
  \bibfield  {author} {\bibinfo {author} {\bibfnamefont {P.~L.}\ \bibnamefont {{Bhatnagar}}}, \bibinfo {author} {\bibfnamefont {E.~P.}\ \bibnamefont {{Gross}}}, \ and\ \bibinfo {author} {\bibfnamefont {M.}~\bibnamefont {{Krook}}},\ }\href {\doibase 10.1103/PhysRev.94.511} {\bibfield  {journal} {\bibinfo  {journal} {Physical Review}\ }\textbf {\bibinfo {volume} {94}},\ \bibinfo {pages} {511} (\bibinfo {year} {1954})}\BibitemShut {NoStop}%
\bibitem [{\citenamefont {Weinberg}(2008)}]{Weinberg:2008zzc}%
  \BibitemOpen
  \bibfield  {author} {\bibinfo {author} {\bibfnamefont {S.}~\bibnamefont {Weinberg}},\ }\href@noop {} {\emph {\bibinfo {title} {{Cosmology}}}}\ (\bibinfo {year} {2008})\BibitemShut {NoStop}%
\bibitem [{Note2()}]{Note2}%
  \BibitemOpen
  \bibinfo {note} {Transient solutions get damped in a timescale $\Gamma _{\protect \rm th}^{-1}\ll H^{-1}$~\cite {LongPaper}.}\BibitemShut {Stop}%
\bibitem [{\citenamefont {Peskin}\ and\ \citenamefont {Schroeder}(1995)}]{Peskin:1995ev}%
  \BibitemOpen
  \bibfield  {author} {\bibinfo {author} {\bibfnamefont {M.~E.}\ \bibnamefont {Peskin}}\ and\ \bibinfo {author} {\bibfnamefont {D.~V.}\ \bibnamefont {Schroeder}},\ }\href {\doibase 10.1201/9780429503559} {\emph {\bibinfo {title} {{An Introduction to quantum field theory}}}}\ (\bibinfo  {publisher} {Addison-Wesley},\ \bibinfo {address} {Reading, USA},\ \bibinfo {year} {1995})\BibitemShut {NoStop}%
\bibitem [{Note3()}]{Note3}%
  \BibitemOpen
  \bibinfo {note} {The opposite regime, $\Phi \lesssim \phi _{\protect \rm th}$, corresponds to what is known as the thermal misalignment scenario \cite {Batell:2021ofv,Batell:2022qvr}.}\BibitemShut {Stop}%
\bibitem [{\citenamefont {Pitrou}\ \emph {et~al.}(2018)\citenamefont {Pitrou}, \citenamefont {Coc}, \citenamefont {Uzan},\ and\ \citenamefont {Vangioni}}]{Pitrou:2018cgg}%
  \BibitemOpen
  \bibfield  {author} {\bibinfo {author} {\bibfnamefont {C.}~\bibnamefont {Pitrou}}, \bibinfo {author} {\bibfnamefont {A.}~\bibnamefont {Coc}}, \bibinfo {author} {\bibfnamefont {J.-P.}\ \bibnamefont {Uzan}}, \ and\ \bibinfo {author} {\bibfnamefont {E.}~\bibnamefont {Vangioni}},\ }\href {\doibase 10.1016/j.physrep.2018.04.005} {\bibfield  {journal} {\bibinfo  {journal} {Phys. Rept.}\ }\textbf {\bibinfo {volume} {754}},\ \bibinfo {pages} {1} (\bibinfo {year} {2018})},\ \Eprint {http://arxiv.org/abs/1801.08023} {arXiv:1801.08023 [astro-ph.CO]} \BibitemShut {NoStop}%
\bibitem [{\citenamefont {Lorenz}\ \emph {et~al.}(2021)\citenamefont {Lorenz}, \citenamefont {Funcke}, \citenamefont {L{\"o}ffler},\ and\ \citenamefont {Calabrese}}]{Lorenz:2021alz}%
  \BibitemOpen
  \bibfield  {author} {\bibinfo {author} {\bibfnamefont {C.~S.}\ \bibnamefont {Lorenz}}, \bibinfo {author} {\bibfnamefont {L.}~\bibnamefont {Funcke}}, \bibinfo {author} {\bibfnamefont {M.}~\bibnamefont {L{\"o}ffler}}, \ and\ \bibinfo {author} {\bibfnamefont {E.}~\bibnamefont {Calabrese}},\ }\href {\doibase 10.1103/PhysRevD.104.123518} {\bibfield  {journal} {\bibinfo  {journal} {Phys. Rev. D}\ }\textbf {\bibinfo {volume} {104}},\ \bibinfo {pages} {123518} (\bibinfo {year} {2021})},\ \Eprint {http://arxiv.org/abs/2102.13618} {arXiv:2102.13618 [astro-ph.CO]} \BibitemShut {NoStop}%
\bibitem [{Note4()}]{Note4}%
  \BibitemOpen
  \bibinfo {note} {The $\DOTSB \sum@ \slimits@ _i M_{\nu _i}$ is also limited to varying degrees at lower temperatures, however these are less constraining in our scenario, which predicts an $M_\nu $ that is decreasing or constant with time.}\BibitemShut {Stop}%
\bibitem [{Note5()}]{Note5}%
  \BibitemOpen
  \bibinfo {note} {We find that in the model under consideration, $\left <\Upsilon \right >_{\protect \rm osc}$ never exceeds Hubble damping in the opposite regime, $\Phi \ll m_\nu /g$. This can be seen as follows. At each temperature $T\lesssim 100\mathinner {\protect \mathrm {GeV}}$, there are couplings $g\gtrsim G_F^2T^4$ that saturates the $\left <\Upsilon \right >_{\protect \rm osc}/H$ at its maximum value $\sim G_F^2m_\nu ^2T\sim 10^{-10}(T/100\mathinner {\protect \mathrm {GeV}})$. Repeating the same argument at $T\gtrsim 100\mathinner {\protect \mathrm {GeV}}$ yields a maximum $\left <\Upsilon \right >_{\protect \rm osc}/H$ that scales as $\propto T^{-3}$. We therefore conclude that in the $\Phi \ll m_\nu /g$ regime, the highest possible $\left <\Upsilon \right >_{\protect \rm osc}/H$ occurs at $T\sim 100\mathinner {\protect \mathrm {GeV}}$ and is exceedingly tiny.}\BibitemShut {Stop}%
\bibitem [{Note6()}]{Note6}%
  \BibitemOpen
  \bibinfo {note} {Since $n_\phi \propto M_\phi \Phi ^2\propto a^{-3}$ is an adiabatic invariant (as long as $|\protect \dot {M}_\phi |/M_\phi ^2\sim M_\phi H/M_\phi ^2\ll 1$), and $M_\phi \propto a^{-1}$ in the thermal-mass dominated case, we must have $\Phi \propto a^{-1}$.}\BibitemShut {Stop}%
\bibitem [{\citenamefont {Tenkanen}(2019)}]{Tenkanen:2019aij}%
  \BibitemOpen
  \bibfield  {author} {\bibinfo {author} {\bibfnamefont {T.}~\bibnamefont {Tenkanen}},\ }\href {\doibase 10.1103/PhysRevLett.123.061302} {\bibfield  {journal} {\bibinfo  {journal} {Phys. Rev. Lett.}\ }\textbf {\bibinfo {volume} {123}},\ \bibinfo {pages} {061302} (\bibinfo {year} {2019})},\ \Eprint {http://arxiv.org/abs/1905.01214} {arXiv:1905.01214 [astro-ph.CO]} \BibitemShut {NoStop}%
\bibitem [{\citenamefont {Kolb}\ and\ \citenamefont {Long}(2024)}]{Kolb:2023ydq}%
  \BibitemOpen
  \bibfield  {author} {\bibinfo {author} {\bibfnamefont {E.~W.}\ \bibnamefont {Kolb}}\ and\ \bibinfo {author} {\bibfnamefont {A.~J.}\ \bibnamefont {Long}},\ }\href {\doibase 10.1103/RevModPhys.96.045005} {\bibfield  {journal} {\bibinfo  {journal} {Rev. Mod. Phys.}\ }\textbf {\bibinfo {volume} {96}},\ \bibinfo {pages} {045005} (\bibinfo {year} {2024})},\ \Eprint {http://arxiv.org/abs/2312.09042} {arXiv:2312.09042 [astro-ph.CO]} \BibitemShut {NoStop}%
\bibitem [{\citenamefont {Graham}\ and\ \citenamefont {Scherlis}(2018)}]{Graham:2018jyp}%
  \BibitemOpen
  \bibfield  {author} {\bibinfo {author} {\bibfnamefont {P.~W.}\ \bibnamefont {Graham}}\ and\ \bibinfo {author} {\bibfnamefont {A.}~\bibnamefont {Scherlis}},\ }\href {\doibase 10.1103/PhysRevD.98.035017} {\bibfield  {journal} {\bibinfo  {journal} {Phys. Rev. D}\ }\textbf {\bibinfo {volume} {98}},\ \bibinfo {pages} {035017} (\bibinfo {year} {2018})},\ \Eprint {http://arxiv.org/abs/1805.07362} {arXiv:1805.07362 [hep-ph]} \BibitemShut {NoStop}%
\bibitem [{\citenamefont {Alonso-{\'A}lvarez}\ and\ \citenamefont {Jaeckel}(2018)}]{Alonso-Alvarez:2018tus}%
  \BibitemOpen
  \bibfield  {author} {\bibinfo {author} {\bibfnamefont {G.}~\bibnamefont {Alonso-{\'A}lvarez}}\ and\ \bibinfo {author} {\bibfnamefont {J.}~\bibnamefont {Jaeckel}},\ }\href {\doibase 10.1088/1475-7516/2018/10/022} {\bibfield  {journal} {\bibinfo  {journal} {JCAP}\ }\textbf {\bibinfo {volume} {10}},\ \bibinfo {pages} {022} (\bibinfo {year} {2018})},\ \Eprint {http://arxiv.org/abs/1807.09785} {arXiv:1807.09785 [hep-ph]} \BibitemShut {NoStop}%
\bibitem [{\citenamefont {Garcia}\ \emph {et~al.}(2023)\citenamefont {Garcia}, \citenamefont {Pierre},\ and\ \citenamefont {Verner}}]{Garcia:2023qab}%
  \BibitemOpen
  \bibfield  {author} {\bibinfo {author} {\bibfnamefont {M.~A.~G.}\ \bibnamefont {Garcia}}, \bibinfo {author} {\bibfnamefont {M.}~\bibnamefont {Pierre}}, \ and\ \bibinfo {author} {\bibfnamefont {S.}~\bibnamefont {Verner}},\ }\href {\doibase 10.1103/PhysRevD.108.115024} {\bibfield  {journal} {\bibinfo  {journal} {Phys. Rev. D}\ }\textbf {\bibinfo {volume} {108}},\ \bibinfo {pages} {115024} (\bibinfo {year} {2023})},\ \Eprint {http://arxiv.org/abs/2305.14446} {arXiv:2305.14446 [hep-ph]} \BibitemShut {NoStop}%
\bibitem [{\citenamefont {Preskill}\ \emph {et~al.}(1983)\citenamefont {Preskill}, \citenamefont {Wise},\ and\ \citenamefont {Wilczek}}]{Preskill:1982cy}%
  \BibitemOpen
  \bibfield  {author} {\bibinfo {author} {\bibfnamefont {J.}~\bibnamefont {Preskill}}, \bibinfo {author} {\bibfnamefont {M.~B.}\ \bibnamefont {Wise}}, \ and\ \bibinfo {author} {\bibfnamefont {F.}~\bibnamefont {Wilczek}},\ }\href {\doibase 10.1016/0370-2693(83)90637-8} {\bibfield  {journal} {\bibinfo  {journal} {Phys. Lett. B}\ }\textbf {\bibinfo {volume} {120}},\ \bibinfo {pages} {127} (\bibinfo {year} {1983})}\BibitemShut {NoStop}%
\bibitem [{\citenamefont {Abbott}\ and\ \citenamefont {Sikivie}(1983)}]{Abbott:1982af}%
  \BibitemOpen
  \bibfield  {author} {\bibinfo {author} {\bibfnamefont {L.~F.}\ \bibnamefont {Abbott}}\ and\ \bibinfo {author} {\bibfnamefont {P.}~\bibnamefont {Sikivie}},\ }\href {\doibase 10.1016/0370-2693(83)90638-X} {\bibfield  {journal} {\bibinfo  {journal} {Phys. Lett. B}\ }\textbf {\bibinfo {volume} {120}},\ \bibinfo {pages} {133} (\bibinfo {year} {1983})}\BibitemShut {NoStop}%
\bibitem [{\citenamefont {Dine}\ and\ \citenamefont {Fischler}(1983)}]{Dine:1982ah}%
  \BibitemOpen
  \bibfield  {author} {\bibinfo {author} {\bibfnamefont {M.}~\bibnamefont {Dine}}\ and\ \bibinfo {author} {\bibfnamefont {W.}~\bibnamefont {Fischler}},\ }\href {\doibase 10.1016/0370-2693(83)90639-1} {\bibfield  {journal} {\bibinfo  {journal} {Phys. Lett. B}\ }\textbf {\bibinfo {volume} {120}},\ \bibinfo {pages} {137} (\bibinfo {year} {1983})}\BibitemShut {NoStop}%
\bibitem [{Note7()}]{Note7}%
  \BibitemOpen
  \bibinfo {note} {In a companion paper \cite {LongPaper}, we consider the possibility that $\phi $ constitutes a fraction of the full DM.}\BibitemShut {Stop}%
\bibitem [{\citenamefont {Aghanim}\ \emph {et~al.}(2020)\citenamefont {Aghanim} \emph {et~al.}}]{Planck:2018vyg}%
  \BibitemOpen
  \bibfield  {author} {\bibinfo {author} {\bibfnamefont {N.}~\bibnamefont {Aghanim}} \emph {et~al.} (\bibinfo {collaboration} {Planck}),\ }\href {\doibase 10.1051/0004-6361/201833910} {\bibfield  {journal} {\bibinfo  {journal} {Astron. Astrophys.}\ }\textbf {\bibinfo {volume} {641}},\ \bibinfo {pages} {A6} (\bibinfo {year} {2020})},\ \bibinfo {note} {[Erratum: Astron.Astrophys. 652, C4 (2021)]},\ \Eprint {http://arxiv.org/abs/1807.06209} {arXiv:1807.06209 [astro-ph.CO]} \BibitemShut {NoStop}%
\bibitem [{\citenamefont {Kopp}\ \emph {et~al.}(2018)\citenamefont {Kopp}, \citenamefont {Skordis}, \citenamefont {Thomas},\ and\ \citenamefont {Ili{\'c}}}]{Kopp:2018zxp}%
  \BibitemOpen
  \bibfield  {author} {\bibinfo {author} {\bibfnamefont {M.}~\bibnamefont {Kopp}}, \bibinfo {author} {\bibfnamefont {C.}~\bibnamefont {Skordis}}, \bibinfo {author} {\bibfnamefont {D.~B.}\ \bibnamefont {Thomas}}, \ and\ \bibinfo {author} {\bibfnamefont {S.}~\bibnamefont {Ili{\'c}}},\ }\href {\doibase 10.1103/PhysRevLett.120.221102} {\bibfield  {journal} {\bibinfo  {journal} {Phys. Rev. Lett.}\ }\textbf {\bibinfo {volume} {120}},\ \bibinfo {pages} {221102} (\bibinfo {year} {2018})},\ \Eprint {http://arxiv.org/abs/1802.09541} {arXiv:1802.09541 [astro-ph.CO]} \BibitemShut {NoStop}%
\bibitem [{\citenamefont {Lambiase}\ \emph {et~al.}(2025)\citenamefont {Lambiase}, \citenamefont {Poddar},\ and\ \citenamefont {Visinelli}}]{Lambiase:2025twn}%
  \BibitemOpen
  \bibfield  {author} {\bibinfo {author} {\bibfnamefont {G.}~\bibnamefont {Lambiase}}, \bibinfo {author} {\bibfnamefont {T.~K.}\ \bibnamefont {Poddar}}, \ and\ \bibinfo {author} {\bibfnamefont {L.}~\bibnamefont {Visinelli}},\ }\href {\doibase 10.1103/yfwv-37ss} {\bibfield  {journal} {\bibinfo  {journal} {Phys. Rev. D}\ }\textbf {\bibinfo {volume} {112}},\ \bibinfo {pages} {016010} (\bibinfo {year} {2025})},\ \Eprint {http://arxiv.org/abs/2503.02940} {arXiv:2503.02940 [hep-ph]} \BibitemShut {NoStop}%
\bibitem [{\citenamefont {Mathur}\ \emph {et~al.}(2020)\citenamefont {Mathur}, \citenamefont {Rajendran},\ and\ \citenamefont {Tanin}}]{Mathur:2020aqv}%
  \BibitemOpen
  \bibfield  {author} {\bibinfo {author} {\bibfnamefont {A.}~\bibnamefont {Mathur}}, \bibinfo {author} {\bibfnamefont {S.}~\bibnamefont {Rajendran}}, \ and\ \bibinfo {author} {\bibfnamefont {E.~H.}\ \bibnamefont {Tanin}},\ }\href {\doibase 10.1103/PhysRevD.102.055015} {\bibfield  {journal} {\bibinfo  {journal} {Phys. Rev. D}\ }\textbf {\bibinfo {volume} {102}},\ \bibinfo {pages} {055015} (\bibinfo {year} {2020})},\ \Eprint {http://arxiv.org/abs/2004.12326} {arXiv:2004.12326 [hep-ph]} \BibitemShut {NoStop}%
\bibitem [{\citenamefont {Fukuda}\ and\ \citenamefont {Nakayama}(2020)}]{Fukuda:2019ewf}%
  \BibitemOpen
  \bibfield  {author} {\bibinfo {author} {\bibfnamefont {H.}~\bibnamefont {Fukuda}}\ and\ \bibinfo {author} {\bibfnamefont {K.}~\bibnamefont {Nakayama}},\ }\href {\doibase 10.1007/JHEP01(2020)128} {\bibfield  {journal} {\bibinfo  {journal} {JHEP}\ }\textbf {\bibinfo {volume} {01}},\ \bibinfo {pages} {128} (\bibinfo {year} {2020})},\ \Eprint {http://arxiv.org/abs/1910.06308} {arXiv:1910.06308 [hep-ph]} \BibitemShut {NoStop}%
\bibitem [{\citenamefont {Arvanitaki}\ \emph {et~al.}(2015)\citenamefont {Arvanitaki}, \citenamefont {Baryakhtar},\ and\ \citenamefont {Huang}}]{Arvanitaki:2014wva}%
  \BibitemOpen
  \bibfield  {author} {\bibinfo {author} {\bibfnamefont {A.}~\bibnamefont {Arvanitaki}}, \bibinfo {author} {\bibfnamefont {M.}~\bibnamefont {Baryakhtar}}, \ and\ \bibinfo {author} {\bibfnamefont {X.}~\bibnamefont {Huang}},\ }\href {\doibase 10.1103/PhysRevD.91.084011} {\bibfield  {journal} {\bibinfo  {journal} {Phys. Rev. D}\ }\textbf {\bibinfo {volume} {91}},\ \bibinfo {pages} {084011} (\bibinfo {year} {2015})},\ \Eprint {http://arxiv.org/abs/1411.2263} {arXiv:1411.2263 [hep-ph]} \BibitemShut {NoStop}%
\bibitem [{Note8()}]{Note8}%
  \BibitemOpen
  \bibinfo {note} {These include but not limited to those considered in our companion paper \cite {LongPaper}.}\BibitemShut {Stop}%
\bibitem [{\citenamefont {Fardon}\ \emph {et~al.}(2004)\citenamefont {Fardon}, \citenamefont {Nelson},\ and\ \citenamefont {Weiner}}]{Fardon:2003eh}%
  \BibitemOpen
  \bibfield  {author} {\bibinfo {author} {\bibfnamefont {R.}~\bibnamefont {Fardon}}, \bibinfo {author} {\bibfnamefont {A.~E.}\ \bibnamefont {Nelson}}, \ and\ \bibinfo {author} {\bibfnamefont {N.}~\bibnamefont {Weiner}},\ }\href {\doibase 10.1088/1475-7516/2004/10/005} {\bibfield  {journal} {\bibinfo  {journal} {JCAP}\ }\textbf {\bibinfo {volume} {10}},\ \bibinfo {pages} {005} (\bibinfo {year} {2004})},\ \Eprint {http://arxiv.org/abs/astro-ph/0309800} {arXiv:astro-ph/0309800} \BibitemShut {NoStop}%
\bibitem [{\citenamefont {Kaplan}\ \emph {et~al.}(2004)\citenamefont {Kaplan}, \citenamefont {Nelson},\ and\ \citenamefont {Weiner}}]{Kaplan:2004dq}%
  \BibitemOpen
  \bibfield  {author} {\bibinfo {author} {\bibfnamefont {D.~B.}\ \bibnamefont {Kaplan}}, \bibinfo {author} {\bibfnamefont {A.~E.}\ \bibnamefont {Nelson}}, \ and\ \bibinfo {author} {\bibfnamefont {N.}~\bibnamefont {Weiner}},\ }\href {\doibase 10.1103/PhysRevLett.93.091801} {\bibfield  {journal} {\bibinfo  {journal} {Phys. Rev. Lett.}\ }\textbf {\bibinfo {volume} {93}},\ \bibinfo {pages} {091801} (\bibinfo {year} {2004})},\ \Eprint {http://arxiv.org/abs/hep-ph/0401099} {arXiv:hep-ph/0401099} \BibitemShut {NoStop}%
\bibitem [{\citenamefont {Sakstein}\ and\ \citenamefont {Trodden}(2020)}]{Sakstein:2019fmf}%
  \BibitemOpen
  \bibfield  {author} {\bibinfo {author} {\bibfnamefont {J.}~\bibnamefont {Sakstein}}\ and\ \bibinfo {author} {\bibfnamefont {M.}~\bibnamefont {Trodden}},\ }\href {\doibase 10.1103/PhysRevLett.124.161301} {\bibfield  {journal} {\bibinfo  {journal} {Phys. Rev. Lett.}\ }\textbf {\bibinfo {volume} {124}},\ \bibinfo {pages} {161301} (\bibinfo {year} {2020})},\ \Eprint {http://arxiv.org/abs/1911.11760} {arXiv:1911.11760 [astro-ph.CO]} \BibitemShut {NoStop}%
\bibitem [{\citenamefont {Brookfield}\ \emph {et~al.}(2006)\citenamefont {Brookfield}, \citenamefont {van~de Bruck}, \citenamefont {Mota},\ and\ \citenamefont {Tocchini-Valentini}}]{Brookfield:2005bz}%
  \BibitemOpen
  \bibfield  {author} {\bibinfo {author} {\bibfnamefont {A.~W.}\ \bibnamefont {Brookfield}}, \bibinfo {author} {\bibfnamefont {C.}~\bibnamefont {van~de Bruck}}, \bibinfo {author} {\bibfnamefont {D.~F.}\ \bibnamefont {Mota}}, \ and\ \bibinfo {author} {\bibfnamefont {D.}~\bibnamefont {Tocchini-Valentini}},\ }\href {\doibase 10.1103/PhysRevD.73.083515} {\bibfield  {journal} {\bibinfo  {journal} {Phys. Rev. D}\ }\textbf {\bibinfo {volume} {73}},\ \bibinfo {pages} {083515} (\bibinfo {year} {2006})},\ \bibinfo {note} {[Erratum: Phys.Rev.D 76, 049901 (2007)]},\ \Eprint {http://arxiv.org/abs/astro-ph/0512367} {arXiv:astro-ph/0512367} \BibitemShut {NoStop}%
\bibitem [{\citenamefont {Kamionkowski}\ and\ \citenamefont {Mathur}(2025)}]{Kamionkowski:2024axz}%
  \BibitemOpen
  \bibfield  {author} {\bibinfo {author} {\bibfnamefont {M.}~\bibnamefont {Kamionkowski}}\ and\ \bibinfo {author} {\bibfnamefont {A.}~\bibnamefont {Mathur}},\ }\href {\doibase 10.1103/PhysRevD.111.063551} {\bibfield  {journal} {\bibinfo  {journal} {Phys. Rev. D}\ }\textbf {\bibinfo {volume} {111}},\ \bibinfo {pages} {063551} (\bibinfo {year} {2025})},\ \Eprint {http://arxiv.org/abs/2411.09747} {arXiv:2411.09747 [hep-ph]} \BibitemShut {NoStop}%
\bibitem [{\citenamefont {Ge}\ and\ \citenamefont {Parke}(2019)}]{Ge:2018uhz}%
  \BibitemOpen
  \bibfield  {author} {\bibinfo {author} {\bibfnamefont {S.-F.}\ \bibnamefont {Ge}}\ and\ \bibinfo {author} {\bibfnamefont {S.~J.}\ \bibnamefont {Parke}},\ }\href {\doibase 10.1103/PhysRevLett.122.211801} {\bibfield  {journal} {\bibinfo  {journal} {Phys. Rev. Lett.}\ }\textbf {\bibinfo {volume} {122}},\ \bibinfo {pages} {211801} (\bibinfo {year} {2019})},\ \Eprint {http://arxiv.org/abs/1812.08376} {arXiv:1812.08376 [hep-ph]} \BibitemShut {NoStop}%
\bibitem [{\citenamefont {Smirnov}\ and\ \citenamefont {Xu}(2019)}]{Smirnov:2019cae}%
  \BibitemOpen
  \bibfield  {author} {\bibinfo {author} {\bibfnamefont {A.~Y.}\ \bibnamefont {Smirnov}}\ and\ \bibinfo {author} {\bibfnamefont {X.-J.}\ \bibnamefont {Xu}},\ }\href {\doibase 10.1007/JHEP12(2019)046} {\bibfield  {journal} {\bibinfo  {journal} {JHEP}\ }\textbf {\bibinfo {volume} {12}},\ \bibinfo {pages} {046} (\bibinfo {year} {2019})},\ \Eprint {http://arxiv.org/abs/1909.07505} {arXiv:1909.07505 [hep-ph]} \BibitemShut {NoStop}%
\bibitem [{\citenamefont {Babu}\ \emph {et~al.}(2020)\citenamefont {Babu}, \citenamefont {Chauhan},\ and\ \citenamefont {Bhupal~Dev}}]{Babu:2019iml}%
  \BibitemOpen
  \bibfield  {author} {\bibinfo {author} {\bibfnamefont {K.~S.}\ \bibnamefont {Babu}}, \bibinfo {author} {\bibfnamefont {G.}~\bibnamefont {Chauhan}}, \ and\ \bibinfo {author} {\bibfnamefont {P.~S.}\ \bibnamefont {Bhupal~Dev}},\ }\href {\doibase 10.1103/PhysRevD.101.095029} {\bibfield  {journal} {\bibinfo  {journal} {Phys. Rev. D}\ }\textbf {\bibinfo {volume} {101}},\ \bibinfo {pages} {095029} (\bibinfo {year} {2020})},\ \Eprint {http://arxiv.org/abs/1912.13488} {arXiv:1912.13488 [hep-ph]} \BibitemShut {NoStop}%
\bibitem [{\citenamefont {Venzor}\ \emph {et~al.}(2021)\citenamefont {Venzor}, \citenamefont {P{\'e}rez-Lorenzana},\ and\ \citenamefont {De-Santiago}}]{Venzor:2020ova}%
  \BibitemOpen
  \bibfield  {author} {\bibinfo {author} {\bibfnamefont {J.}~\bibnamefont {Venzor}}, \bibinfo {author} {\bibfnamefont {A.}~\bibnamefont {P{\'e}rez-Lorenzana}}, \ and\ \bibinfo {author} {\bibfnamefont {J.}~\bibnamefont {De-Santiago}},\ }\href {\doibase 10.1103/PhysRevD.103.043534} {\bibfield  {journal} {\bibinfo  {journal} {Phys. Rev. D}\ }\textbf {\bibinfo {volume} {103}},\ \bibinfo {pages} {043534} (\bibinfo {year} {2021})},\ \Eprint {http://arxiv.org/abs/2009.08104} {arXiv:2009.08104 [hep-ph]} \BibitemShut {NoStop}%
\bibitem [{\citenamefont {Dutta}\ \emph {et~al.}(2023)\citenamefont {Dutta}, \citenamefont {Ghosh}, \citenamefont {Li}, \citenamefont {Thompson},\ and\ \citenamefont {Verma}}]{Dutta:2022fdt}%
  \BibitemOpen
  \bibfield  {author} {\bibinfo {author} {\bibfnamefont {B.}~\bibnamefont {Dutta}}, \bibinfo {author} {\bibfnamefont {S.}~\bibnamefont {Ghosh}}, \bibinfo {author} {\bibfnamefont {T.}~\bibnamefont {Li}}, \bibinfo {author} {\bibfnamefont {A.}~\bibnamefont {Thompson}}, \ and\ \bibinfo {author} {\bibfnamefont {A.}~\bibnamefont {Verma}},\ }\href {\doibase 10.1007/JHEP03(2023)163} {\bibfield  {journal} {\bibinfo  {journal} {JHEP}\ }\textbf {\bibinfo {volume} {03}},\ \bibinfo {pages} {163} (\bibinfo {year} {2023})},\ \Eprint {http://arxiv.org/abs/2209.13566} {arXiv:2209.13566 [hep-ph]} \BibitemShut {NoStop}%
\bibitem [{\citenamefont {Berryman}\ \emph {et~al.}(2023)\citenamefont {Berryman} \emph {et~al.}}]{Berryman:2022hds}%
  \BibitemOpen
  \bibfield  {author} {\bibinfo {author} {\bibfnamefont {J.~M.}\ \bibnamefont {Berryman}} \emph {et~al.},\ }\href {\doibase 10.1016/j.dark.2023.101267} {\bibfield  {journal} {\bibinfo  {journal} {Phys. Dark Univ.}\ }\textbf {\bibinfo {volume} {42}},\ \bibinfo {pages} {101267} (\bibinfo {year} {2023})},\ \Eprint {http://arxiv.org/abs/2203.01955} {arXiv:2203.01955 [hep-ph]} \BibitemShut {NoStop}%
\bibitem [{\citenamefont {Kaplan}\ \emph {et~al.}(2025)\citenamefont {Kaplan}, \citenamefont {Luo},\ and\ \citenamefont {Rajendran}}]{Kaplan:2024ydw}%
  \BibitemOpen
  \bibfield  {author} {\bibinfo {author} {\bibfnamefont {D.~E.}\ \bibnamefont {Kaplan}}, \bibinfo {author} {\bibfnamefont {X.}~\bibnamefont {Luo}}, \ and\ \bibinfo {author} {\bibfnamefont {S.}~\bibnamefont {Rajendran}},\ }\href {\doibase 10.1103/PhysRevD.111.055019} {\bibfield  {journal} {\bibinfo  {journal} {Phys. Rev. D}\ }\textbf {\bibinfo {volume} {111}},\ \bibinfo {pages} {055019} (\bibinfo {year} {2025})},\ \Eprint {http://arxiv.org/abs/2412.20766} {arXiv:2412.20766 [hep-ph]} \BibitemShut {NoStop}%
\bibitem [{\citenamefont {Green}\ \emph {et~al.}(2021)\citenamefont {Green}, \citenamefont {Kaplan},\ and\ \citenamefont {Rajendran}}]{Green:2021gdc}%
  \BibitemOpen
  \bibfield  {author} {\bibinfo {author} {\bibfnamefont {D.}~\bibnamefont {Green}}, \bibinfo {author} {\bibfnamefont {D.~E.}\ \bibnamefont {Kaplan}}, \ and\ \bibinfo {author} {\bibfnamefont {S.}~\bibnamefont {Rajendran}},\ }\href {\doibase 10.1007/JHEP11(2021)162} {\bibfield  {journal} {\bibinfo  {journal} {JHEP}\ }\textbf {\bibinfo {volume} {11}},\ \bibinfo {pages} {162} (\bibinfo {year} {2021})},\ \Eprint {http://arxiv.org/abs/2108.06928} {arXiv:2108.06928 [hep-ph]} \BibitemShut {NoStop}%
\bibitem [{\citenamefont {Craig}\ \emph {et~al.}(2024)\citenamefont {Craig}, \citenamefont {Green}, \citenamefont {Meyers},\ and\ \citenamefont {Rajendran}}]{Craig:2024tky}%
  \BibitemOpen
  \bibfield  {author} {\bibinfo {author} {\bibfnamefont {N.}~\bibnamefont {Craig}}, \bibinfo {author} {\bibfnamefont {D.}~\bibnamefont {Green}}, \bibinfo {author} {\bibfnamefont {J.}~\bibnamefont {Meyers}}, \ and\ \bibinfo {author} {\bibfnamefont {S.}~\bibnamefont {Rajendran}},\ }\href {\doibase 10.1007/JHEP09(2024)097} {\bibfield  {journal} {\bibinfo  {journal} {JHEP}\ }\textbf {\bibinfo {volume} {09}},\ \bibinfo {pages} {097} (\bibinfo {year} {2024})},\ \Eprint {http://arxiv.org/abs/2405.00836} {arXiv:2405.00836 [astro-ph.CO]} \BibitemShut {NoStop}%
\bibitem [{\citenamefont {Esteban}\ and\ \citenamefont {Salvado}(2021)}]{Esteban:2021ozz}%
  \BibitemOpen
  \bibfield  {author} {\bibinfo {author} {\bibfnamefont {I.}~\bibnamefont {Esteban}}\ and\ \bibinfo {author} {\bibfnamefont {J.}~\bibnamefont {Salvado}},\ }\href {\doibase 10.1088/1475-7516/2021/05/036} {\bibfield  {journal} {\bibinfo  {journal} {JCAP}\ }\textbf {\bibinfo {volume} {05}},\ \bibinfo {pages} {036} (\bibinfo {year} {2021})},\ \Eprint {http://arxiv.org/abs/2101.05804} {arXiv:2101.05804 [hep-ph]} \BibitemShut {NoStop}%
\bibitem [{\citenamefont {Wang}\ \emph {et~al.}(2025)\citenamefont {Wang}, \citenamefont {Xu},\ and\ \citenamefont {Zhou}}]{Wang:2025qap}%
  \BibitemOpen
  \bibfield  {author} {\bibinfo {author} {\bibfnamefont {I.~R.}\ \bibnamefont {Wang}}, \bibinfo {author} {\bibfnamefont {X.-J.}\ \bibnamefont {Xu}}, \ and\ \bibinfo {author} {\bibfnamefont {B.}~\bibnamefont {Zhou}},\ }\href@noop {} {\  (\bibinfo {year} {2025})},\ \Eprint {http://arxiv.org/abs/2501.07624} {arXiv:2501.07624 [hep-ph]} \BibitemShut {NoStop}%
\bibitem [{\citenamefont {Kofman}\ \emph {et~al.}(1997)\citenamefont {Kofman}, \citenamefont {Linde},\ and\ \citenamefont {Starobinsky}}]{Kofman:1997yn}%
  \BibitemOpen
  \bibfield  {author} {\bibinfo {author} {\bibfnamefont {L.}~\bibnamefont {Kofman}}, \bibinfo {author} {\bibfnamefont {A.~D.}\ \bibnamefont {Linde}}, \ and\ \bibinfo {author} {\bibfnamefont {A.~A.}\ \bibnamefont {Starobinsky}},\ }\href {\doibase 10.1103/PhysRevD.56.3258} {\bibfield  {journal} {\bibinfo  {journal} {Phys. Rev. D}\ }\textbf {\bibinfo {volume} {56}},\ \bibinfo {pages} {3258} (\bibinfo {year} {1997})},\ \Eprint {http://arxiv.org/abs/hep-ph/9704452} {arXiv:hep-ph/9704452} \BibitemShut {NoStop}%
\bibitem [{\citenamefont {Banerjee}\ \emph {et~al.}(pear)\citenamefont {Banerjee}, \citenamefont {Nguyen},\ and\ \citenamefont {Tanin}}]{LongPaper}%
  \BibitemOpen
  \bibfield  {author} {\bibinfo {author} {\bibfnamefont {A.}~\bibnamefont {Banerjee}}, \bibinfo {author} {\bibfnamefont {N.~H.}\ \bibnamefont {Nguyen}}, \ and\ \bibinfo {author} {\bibfnamefont {E.~H.}\ \bibnamefont {Tanin}},\ }\href@noop {} {\  (\bibinfo {year} {to appear})}\BibitemShut {NoStop}%
\end{thebibliography}%

\end{document}